\documentclass[12pt,makeidx]{amsbook}
  \usepackage{amssymb}
   \usepackage{graphics}
    \usepackage[dvips]{epsfig}
   \usepackage{times}
   \usepackage{tikz}
   \usepackage{tcolorbox}
   \usepackage{mathtools}
\labelformat{section}{\thechapter.\thesection}
\labelformat{subsection}{\thechapter.\thesubsection}
   \makeindex
   \newtheorem{theorem}{Theorem}[chapter]
   \newtheorem{lemma}{Lemma}[chapter]
   \newtheorem{corollary}{Corollary}[chapter]
   \newtheorem{proposition}{Proposition}[chapter]
   \newtheorem{ex}{Ex.}[chapter]
   \newtheorem{remark}{Remark}[chapter]

   \newcommand{\D}{\displaystyle}
   
   \def\C{{\mathbb C}}
   
   \def\H{{\mathbb H}}

   \def\R{{\mathbb R}}

   \def\Pf{{\it Proof.$\;\;$}}

   \def\qed{\hspace{12cm}$\diamond$}

    \def\core{\mbox{\rm core}}

   \def\tr{\mbox{\rm tr~}}

   \def\({\langle}
   \def\){\rangle}
   \def\mb{\boldsymbol}

   \def\im{{\rm i}}

   \def\cF{{\mathcal F}}

   \def\cN{{\mathcal N}}
   \def\cO{{\mathcal O}}
   \def\cP{{\mathcal P}}

   \def\cS{{\mathcal S}}

   \def\cX{{\mathcal X}}
   \def\cY{{\mathcal Y}}

   \def\1{\mb1}

   \def\v0{{\bf 0}}

   \def\vx{{\bf x}}
   \def\vy{{\bf y}}

   \def\ov{\overline}

\begin{document}
\frenchspacing
\frontmatter
\begin{titlepage}

{} \vspace{3cm}

\bigskip
\begin{center}
{\bf\Large Mathematical Game Theory}
\end{center}

\vspace{1cm}
 \centerline{\bf\Large Ulrich Faigle}

\vspace{4cm}

\vspace{6cm}
\begin{center}
{\large Department of Mathematics\\
University of Cologne\\
Cologne, Germany\\

\medskip
\normalsize
 faigle@zpr.uni-koeln.de}
\end{center}
\end{titlepage}

\tableofcontents
\chapter*{Preface}

People have gambled and played games for thousands of years. Yet, only in the 17th century we see
a  serious attempt for a scientific approach to the subject. The combinatorial foundations of probability
theory were developed by various mathematicians such as \textsc{J. Bernoulli}\footnote{\textsc{J.
Bernoulli (1654-1705)}}~\cite{Bernoulli1713} as a means to understand games of chance (mostly involving rolls of dice) and to make conjectures according to mathematical principles.

\medskip
Since then, game theory has grown into a wide field and appears at times quite removed from its
combinatorial roots. The notion of a \emph{game} has been broadened to encompass all kinds of human
behavior and interactions of individuals or of groups and societies (see, \emph{e.g.},
\textsc{Berne}~\cite{Berne1964}). Much of current research studies humans in economic and social contexts
and seeks to discover behavioral laws in analogy to physical laws.

\medskip
The role of mathematics in this endeavor, however, has been quite limited so far. One major reason lies
certainly in the fact that players in real life often behave differently than a simple mathematical model
would predict. So seemingly paradoxical situations exist where people appear to contradict the
straightforward analysis of the mathematical model builder. A famous such example is the \emph{chain store
paradox} of \textsc{Selten}\footnote{\textsc{R. Selten (1930-2016)}}~\cite{Selten1978}.

\medskip
This is not withstanding the ground breaking work of \textsc{von Neumann} and
\textsc{Morgenstern}\footnote{\textsc{J. von Neumann (1903-1953), O. Morgenstern
(1902-1977)}}~\cite{vNeumann-Morgenstern}, who have proposed an axiomatic approach to notions of utilities
and rational behavior of the players of a game.

\medskip
As interesting and worthwhile as research into laws that govern psycho\-logical, social or economic
behavior of humans may be, the present \emph{Mathematical Game Theory} is not about these aspects of game
theory. In the center of our attention are mathematical models that may be useful for the ana\-lysis of
game-theoretic situations. We are concerned with the mathe\-matics of game-theoretic models but leave the
question aside whether a particular model describes a particular situation in real life appropriately.

\medskip
The mathematical analysis of a game-theoretic model treats objects neutrally. Elements and sets have no
feelings {\it per se} and show no psychological behavior. They are neither generous nor cost conscious
unless such features are built into the model as clearly formulated mathematical properties. The advantage
of mathematical neutrality is substantial, however, because it allows us to embed the mathematical
analysis into a much wider framework.

\medskip
This introduction into mathematical game theory sees games being played on (possibly quite general)
\emph{systems}. A move of a game then correspond to a transition of a system from one state to another.
Such an  approach reveals a close connection with fundamental physical systems {\it via} the same
underlying mathematics. Indeed, it is hoped that mathematical game theory may eventually play a role for
real world games akin to the role of theoretical physics to real world physical systems.

\medskip
The reader of this introductory text is expected to have knowledge in mathematics, perhaps at the
level of a first course in linear algebra and real analysis. Nevertheless, the text will review relevant mathematical notions and properties and point to the literature for further details.

\medskip
The reader is furthermore expected to read the text ''actively'. ''Ex.'' marks not only an ''example'' but also an
''exercise'' that might deepen the understanding of the mathematical development.

\medskip
The book is based on a one-term course on the subject the author has presented repeatedly at the University of
Cologne to pre-master and master level students with an interest in applied mathematics, operations research and
mathe\-matical modelling.

\medskip
It is dedicated to the memory of \textsc{Walter Kern}\footnote{\textsc{W. Kern} (1957-2021)}.

\mainmatter
\setcounter{chapter}{0}

\part{Introduction}

\chapter{Mathematical Models of the Real World}\label{chap:Mathematical-Models}

\begin{tcolorbox}
{\small This introductory chapter discusses mathematical models, sketches the mathematical tools for their
analysis, defines systems in general and systems of decisions in particular. Games are introduced
from a general point of view and it is indicated how they may arise in combinatorial, economic, social,
physical and other contexts.}
\end{tcolorbox}

\section{Mathematical modelling}\label{sec:mathematical-modelling}
Mathematics is \emph{the} powerful human instrument to analyze and to structure observations and to
possibly  discover natural ''laws''. These laws  are logical principles that allow us not only to
understand observed phenomena ({\emph{i.e.}, the so-called {\em real world})\index{real world}  but also
to compute possible evolutions of current situations and thus to attempt a ''look into the future''.

\medskip
Why is that so? An answer to this question is difficult if not impossible. There is a wide-spread belief
that \emph{mathematics is the language of the universe}\footnote{\textsc{Galileo Galilei} (1564-1642)}. So
everything can supposedly be captured by mathematics and all mathematical de\-ductions reveal facts about
the real world. I do not know whether this is true. But even if it were, one would have to be careful with
real-world interpretations of mathe\-matics, nonetheless. A simple example may illustrate the difficulty:

\medskip
While apples on a tree are counted in terms of natural numbers,  it would certainly be erroneous to conclude:
\emph{for every natural number $n$, there exists a tree with $n$ apples}. In other words, when we use the
set of nonnegative integers to describe the number of apples on a tree, our mathematical model will comprise
mathematical objects that have no real counterparts.

\medskip
Theoretically, one could try to get out of the apple dilemma by restricting the mathe\-matical model to
those numbers $n$ that are realized by apple trees. But such a restricted model would be of no practical
use as neither the set of such apple numbers $n$ nor its specific algebraic structure is explicitly known.
Indeed, while the sum $m+n$ of two natural numbers $m$ and $n$ is a natural number, it is not clear
whether the existence of two apple trees with $m$ resp. $n$ apples guarantees the existence of an apple
tree with $m+n$ apples.

\medskip
In general, a mathematical model of a real-world situation is, alas,  not necessarily guaranteed to be
absolutely comprehensive. Mathematical con\-clusions are possibly only theoretical and may suggest objects
and situations which do not exist in reality. One always has to double-check real-world interpretations of
mathe\-matical deductions and ask whether an interpretation is ''reasonable'' in the sense that it is
commensurate with one's own personal experience.

\medskip
In the analysis of a game-theoretic situation, for example, one may want to take the psychology of
individual players into account. A mathematical model of psychological behavior, however, is typically
based on assumptions whose accuracy is unclear. Consequently, mathematically established results within
such models must be interpreted with care, of course.

\medskip
Moreover, similar to physical systems with a large number of particles (like mole\-cules {\it etc.}),
game-theoretic systems with many agents (\emph{e.g.},  traffic systems and economies) are too complex to
analyze by following each of the many agents individually. Hence a practical approach will have to
concentrate on ''group behavior'' and consider statistical  parameters that average over individual
numerical attributes.

\medskip
Having cautioned the reader about the real-world interpretation of mathematical deductions, we will
concentrate on mathematical models (and their mathematics) and leave the interpretation to the reader. Our
emphasis is on \emph{game-theoretic} models. So we should explain what we understand by this.

\medskip
A \emph{game} involves \emph{players} that perform actions which make a given system go through a sequence
of states. When the game ends, the system is in a state according to which the players receive
\emph{rewards}  (or are charged with \emph{costs} or whatever). Many game theorists think of a ''player''
as a humanoid, \emph{i.e.}, a  creature with human feelings, wishes and desires, and thus give it a human
name\footnote{\emph{Alice} and \emph{Bob} are quite popular choices}.

\medskip
Elements of a \emph{mathematical} model, however, do not have humanoid feelings {\it per se}. If they are
to represent objects with wishes and desires, these wishes and desires must be explicitly formulated as
mathematical optimization challenges with specified objective functions and restrictions. Therefore, we
will try to be neutral and refer to ''players'' often just as \emph{agents} with no specified sexual
attributes. In particular, an agent will typically be an ''it'' rather than a ''he'' or a ''she''.

\medskip
This terminological neutrality makes it clear that mathematical game theory comprises many more models
than just those with human players. As we will see, many models of games, decisions, economics and social
sciences have the same underlying mathematics as models of physics and informatics.

\medskip
{\bf Note on continuous and differentiable functions.} Real world phenomena are often modelled with
continuous or even differentiable functions. However,

\medskip
\begin{tcolorbox}
\begin{itemize}
\item {\em There exists no practically feasible test for the continuity\\ or the differentiability of a function}!
\end{itemize}
\end{tcolorbox}

\medskip
Continuity and differentiability, therefore, are \emph{assumptions} of the model builder. These
assumptions appear often reasonable and produce good results in applications. Moreover, they
facilitate the mathematical ana\-lysis. Yet, their appropriateness cannot be proven by tests and experiments. The reader should be aware of the difference between
a mathe\-matical model and its physical origin.

\medskip
{\bf Note on algorithms and computational complexity.} Game-theoretic questions naturally call for mathematical computations within appropriate models. This will become clear in the present text, which also tries to exhibit important links to mathematical optimization theory. However, here is not the place to discuss specific mathe\-matical optimization procedures {\it per se.} There is an abundance of classical mathematical literature on the latter, which can be consulted by the interested reader.

\medskip
The question of the complexity of computations within particular game-theoretic models has attracted the interest of theoretical computer science and created the field of \emph{algorithmic game theory}\footnote{see, \emph{e.g.}, \textsc{Nisan} {\it al.}~\cite{Nisan-et-al2007}}, whose details would exceed the frame and aim of this text and are, therefore, not addressed either.

\section{Mathematical preliminaries}\label{sec:mathematical-preliminaries}
The reader is assumed to have basic mathematical knowledge (at least at the level of an introductory
course on linear algebra). Nevertheless, it is useful to review some of the mathematical terminology.
Further basic facts are outlined in the Appendix.

\subsection{Functions and data representation}\label{sec:functions-data-representation}

\subsection{Algebra of functions and matrices}\label{sec:matrix-algebra}

\subsection{Numbers and algebra}\label{sec:numbers}

\section{Systems}\label{sec:systems}

\section{Games}\label{sec:games}

\part{2-Person-Games}

\chapter{Combinatorial Games}\label{chap:Combinatorial-games}\index{game!combinatorial}

\begin{tcolorbox}
{\small Games can always be understood as to involve two players that execute moves alternatingly. This aspect reveals a recursive character of games. The chapter takes a look at games that are guaranteed to end after a finite number of moves. Finite games are said to be combinatorial. Under the normal winning rule, combinatorial games have an algebraic structure and behave like generalized numbers. Game algebra allows one to explicitly compute winning strategies for nim games, for example.}
\end{tcolorbox}

\section{Alternating players}\label{sec:alternating-players}

\section{Recursiveness}\label{sec:recursiveness}\index{recursive}

\section{Combinatorial games}\label{sec:combinatorial-games}

\section{Winning strategies}\label{sec:winning-strategies}

\section{Algebra of games}\label{sec:algebra-of-games}\index{game!algebra}

\subsection{Congruent games}\label{sec:congruent-games}\index{game!congruent}

\section{Impartial games}\label{sec:impartial-games}

\subsection{Sums of \textsc{Grundy} numbers}\label{sec:Grundy-sums}

\chapter{Zero-sum Games}\label{chap:zero-sum-games}\index{game!zero-sum}

\begin{tcolorbox}
{\small Zero-sum games abstract the model of combinatorial games. Fundamental examples arise naturally as
\textsc{Lagrange} games from mathematical optimization problems and thus furnish an important link between
game theory and mathematical optimization theory. In particular, strategic equilibria in such games
correspond to optimal solutions of optimization problems. Conversely, mathematical optimization techniques
are important tools for the analysis of game-theoretic situations. }
\end{tcolorbox}

\section{Matrix games}\label{sec:matrix-games}

\section{Equilibria}\label{sec:zero-equilibria}\index{equilibrium}

\section{Convex zero-sum games}\label{sec:convex-zero-sum-games}\index{game!convex}

\subsection{Computational aspects}\label{sec:computational-aspects-matrix-games} 

\section{\textsc{Lagrange} games}\label{sec:Lagrange-games}

\subsection{Complementary slackness}\label{sec:complementary-slackness} 

\subsection{The KKT-conditions}\label{sec:KKT-conditions}

\subsection{Shadow prices}\label{sec:shadow-price}\index{shadow price}

\subsection{ Equilibria of convex \textsc{Lagrange} games}\label{sec:convex-Lagrange-equilibria}

\subsection{Linear programs}\label{sec:linear-programs} \index{linear program}  A \emph{linear program} (LP) in \emph{standard form} is an optimization problem of the form
\begin{equation}\label{eq.standardLP}
\max_{x\in \R^n_+}~ c^Tx  \quad\mbox{s.t.}\quad Ax \leq b,
\end{equation}
where $c\in \R^n$ and $b\in \R^m$ are parameter vectors and $A\in \R^{m\times n}$ a matrix, and thus is a mathematical optimization problem with a linear objective function $f(x) = c^Tx$ and restriction function $g(x) = b-Ax$ so that
$$
   g(x)\geq 0 \quad \longleftrightarrow \quad Ax\leq b.
$$
The feasibility region $\cF$ of (\ref{eq.standardLP}) is the set  of all non\-negative solutions of the linear inequality system $Ax\leq b$:
$$
  \cF = P_+(A,b) = \{x\in \R^n \mid Ax\leq b, x\geq 0\}.
$$
The \textsc{Lagrange} function is
\begin{eqnarray*}
L(x,y) &=& c^Tx + y^T(b-Ax) = y^Tb + (c^T -y^TA)x
\end{eqnarray*}
and yields for any $x\geq 0$ and $y\geq 0$:

\medskip
\begin{tcolorbox}
\begin{eqnarray*}
L_1(x) &=& \min_{y\geq 0}~L(x,y) = \left\{\begin{array}{cl} c^Tx &\mbox{if $Ax\leq b$}\\
                   -\infty &\mbox{if $Ax\not\leq b$.}\end{array}\right. \\
L_2(y) &=& \max_{x\geq 0}~L(x,y) = \left\{\begin{array}{cl} b^Ty &\mbox{if $y^TA \geq c^T$}\\
                   +\infty &\mbox{if $y^T A \not\geq  c^T$.}\end{array}\right.
\end{eqnarray*}
\end{tcolorbox}

\medskip\index{dual linear program}\index{linear program!dual}
The optimum value of $L_2$ is found by solving the \emph{dual} associated linear program
\begin{equation}\label{eq.dualstandardLP}
  \min_{y\geq 0}~ L_2(y) = \min_{y\geq 0}~ y^Tb \quad\mbox{s.t.}\quad y^TA \geq c^T.
\end{equation}

\medskip\index{standard linear program}
\begin{ex}\label{ex.dualstandardLP} Since $y^Tb = b^Ty$ and $(c^T-y^TA)x = x^T(c-A^Ty)$ holds, the dual linear program (\ref{eq.dualstandardLP}) can be formulated equivalently in standard form:
\begin{equation}\label{eq.dualstandardcanonicalLP}
\max_{y\in \R^m_+}~ (-b)^Ty \quad\mbox{s.t.}\quad (-A^T)y \leq -c.
\end{equation}
\end{ex}

\medskip
The main theorem on linear programming is:

\medskip
\begin{tcolorbox}
\begin{theorem}[Main LP-Theorem]\label{t.main-LP} For the LP (\ref{eq.standardLP}) the following holds:
\begin{enumerate}
\item[(A)] An optimal solution $x^*$ exists if and only if both the LP (\ref{eq.standardLP}) and the dual LP (\ref{eq.dualstandardLP}) have feasible solutions.
\item[(B)] A feasible $x^*$ is an optimal solution if and only if there exists a dually feasible solution $y^*$ such that
$$
   c^Tx^* = L_1(x^*) = L_2(y^*) = b^Ty^*.
$$
\end{enumerate}
\end{theorem}
\end{tcolorbox}

\Pf  Assume that (\ref{eq.standardLP}) has an optimal solution $x^*$ with value $z^*=c^Tx^*$. Then $c^Tx\leq z^*$ holds for all feasible solutions $x$. So the \textsc{Farkas} Lemma\footnote{see Lemma~\ref{al.Farkas+} in the Appendix}  guarantees the existence of some $y^*\geq 0$ such that
$$
(y^*)^TA \geq c^T \quad\mbox{and}\quad (y^*)^Tb  \leq z^*.
$$
Noticing that  $y^*$ is dually feasible and that $L_1(x^*) \leq L_2(y)$ holds for all $y\geq 0$, we conclude that $y^*$ is, in fact, an optimal dual solution:
$$
  L_2(y^*) = (y^*)^Tb \leq z^* = L_1(x^*) \leq L_2(y^*)\quad\Longrightarrow\quad L_1(x^*) = L_2(y^*).
$$
This argument establishes property (B) and shows that the existence of an optimal solution necessitates the existence a dually feasible solution. Assuming that (\ref{eq.standardLP}) has at least one feasible solution $x$,
it therefore remains to show that the existence a dual feasible solution $y$ implies the existence of an optimal solution.

\medskip
To see this, note first
$$
   w^* = \inf_{y\geq 0}~b^Ty \;\geq\; L_1(x) > -\infty.
$$
So each dually feasible $y$ satisfies $-b^Ty\leq -w^*$. Applying now the \textsc{Farkas} Lemma to the dual linear program  in the form (\ref{eq.dualstandardcanonicalLP}), we find that a  parameter vector $x^*\geq 0$ exists with the property
$$
   Ax^*\leq b \quad\mbox{and}\quad L_1(x^*) \geq w^*.
$$
On the other hand, the primal-dual inequality yields $L_1(x^*)\leq w^*$. So $x^*$ must be an optimal feasible solution.

\qed

\medskip\index{canonical linear program}
\paragraph{\bf General linear programs.} In general, a \emph{linear program} refers to the problem of optimizing a linear objective function over a \emph{polyhedron}\index{polyhedron}, namely the set of solutions of a finite system of linear equalities and inequalities and, therefore, can be formulated as
\begin{equation}\label{eq.general-LP}
\max_{x\in \R^n}~ c^Tx \quad\mbox{s.t.}\quad Ax\leq b, Bx = d
\end{equation}
with coefficient vectors $c\in \R^n$, $b\in \R^m$, $d\in \R^k$ and matrices $A\in \R^{m\times n}$ and $B\in \R^{k\times n}$.

\medskip
If no equalities occur in the formulation (\ref{eq.general-LP}), one has as a linear program in \emph{canonical form}:
\begin{equation}\label{eq.canonical-LP}
  \max_{x\in \R^n}~c^Tx \quad\mbox{s.t.}\quad Ax\leq b.
\end{equation}

Because of the equivalence
$$
    Bx =d \quad \Longleftrightarrow \quad \mbox{$Bx\leq d$ and $-B\leq -d$},
$$
the optimization problem (\ref{eq.general-LP}) can be presented in canonical form:
$$
\max_{x\in \R^n}~c^Tx\quad\mbox{s.t.}\quad \begin{bmatrix} A\\ B\\-B\end{bmatrix} x\leq \begin{pmatrix} b\\  d\\-d\end{pmatrix}.
$$

\medskip
Moreover, since any vector $x\in \R^n$ can be expressed as the difference
$$
x = x^+-x^-
$$
of two (nonnegative) vectors $x^+,x^-\in \R^n_+$, one sees that each linear program in canonical form is equivalent to a linear program in standard form:
$$
\max_{x^+,x^-\geq 0}~ c^Tx^- - c^Tx^- \quad\mbox{s.t.}\quad Ax^+ -Ax^- \leq b.
$$

\medskip
The \textsc{Lagrange} function of the canonical form is the same as for the standard form. Since the domain of $L_1$ is now $X = \R^n$, the utility function $L_2(y)$ differs accordingly:
\begin{tcolorbox}
\begin{eqnarray*}
L_1(x) &=& \min_{y\geq 0}~L(x,y) = \left\{\begin{array}{cl} c^Tx &\mbox{if $Ax\leq b$}\\
                   -\infty &\mbox{if $Ax\not\leq b$.}\end{array}\right. \\
L_2(y) &=& \max_{x\geq 0}~L(x,y) = \left\{\begin{array}{cl} b^Ty &\mbox{if $y^TA =c^T$}\\
                   +\infty &\mbox{if $y^T A \neq  c^T$.}\end{array}\right.
\end{eqnarray*}
\end{tcolorbox}

\medskip\index{dual linear program}\index{linear program}
Relative to the canonical form, the optimum value of $L_2$ is found by solving the linear program
\begin{equation}\label{eq.dualcanonicalLP}
  \min_{y\geq 0}~ L_2(y) = \min_{y\geq 0}~ y^Tb \quad\mbox{s.t.}\quad y^TA = c^T.
\end{equation}

\medskip
Nevertheless, it is straightforward to check that Theorem~\ref{t.main-LP} is literally valid also for a linear program in its canonical form.

\medskip
Linear programming problems are particularly important in applications because they can be solved
efficiently.  In the theory of \emph{cooperative
games} with possibly more than two players (see Chapter~\ref{chap:Cooperative-games}), linear programming
is a structurally analytical tool. We do not go into algorithmic details here but refer to the standard mathematical optimization literature\footnote{\emph{e.g.}, \textsc{Faigle} {\it et al.}~\cite{FaigleKernStill2002}}.

\subsection{Linear programming games}\label{sec:LP-games}\index{games!linear programming}

\chapter{Investing and Betting}\label{chap:Investing-betting}

\begin{tcolorbox}
{\small The opponent of a gambler is usually a player with no specific optimization goal. The opponent's
strategy choices  seem to be determined by chance. Therefore, the gambler will have to decide on strategies with good expected returns. Information plays an important role in the quest for the best decision. Hence the
problem how to model information exchange and common knowledge among (possibly more than two) players
deserves to be addressed as well.}
\end{tcolorbox}

\section{Proportional investing}\label{sec:proportional-investing}

\subsection{Expected utility} 

\subsection{The fortune formula}\label{sec:fortune-formula}

\section{Fair odds}\label{sec:fair-odds}

\section{Betting on alternatives}\label{sec:betting-alternatives}

\section{Betting and information}\label{sec:betting-information}

\section{Common knowledge}\label{sec:common-knowledge}

\subsection{Red and white hats}\label{sec:red-white-hats}
Imagine the following situation:
\begin{itemize}\index{red hats}
\item[(I)] Three girls, $G_1$, $G_2$ and $G_3$, with \emph{red} hats sit in a circle.
\item[(II)] Each girl knows that their hats are either \emph{red} or \emph{white}.
\item[(III)] Each girl can see the color of all hats except her own.
\end{itemize}

\medskip
Now the teacher comes and announces:
\begin{enumerate}
\item \emph{There is at least one red hat.}
\item \emph{I will start counting slowly. As soon as someone knows the color of her hat, she should
    raise her hand.}
\end{enumerate}

\medskip
What will happen? Does the teacher provide information that goes beyond the common knowledge the girls
already have? After all, each girl sees two red  hats -- and hence \emph{knows} that each of the other
girls sees at least one red had as well.

\medskip
Because of (III), the girls know their hat universe $\mathfrak H$ is in one of the $8$ states of possible
color distributions:
$$
\begin{array}{c|cccc|ccc|c}
  &\sigma_1 &\sigma_2 &\sigma_3 &\sigma_4 &\sigma_5 &\sigma_6 &\sigma_7 &\sigma_8 \\ \hline
G_1     &R     &R         &R         &W       &R          &W        &W        &W  \\
G_2   &R     &R         &W         &R       &W          &R        &W        &W   \\
G_3   &R     &W         &R         &R        &W         &W       &R        &W     \\
\end{array}
$$
None of these states can be jointly ruled out. The entropy $H^0_2$ of their common knowledge is:
$$
               H^0_2 = \log_2 8 = 3.
$$
The teacher's announcement, however, rules out the state $\sigma_8$ and reduces the entropy to
$$
               H^1_2 = \log_2 7 < H^0_2,
$$
which means that the teacher has supplied proper additional information.

\medskip
At the teacher's first count, no girl can be sure about her own hat because none sees \emph{two} white
hats. So no hand is raised, which rules out the states $\sigma_5,\sigma_6$ and $\sigma_7$ as
possibilities.

\medskip
Denote now by $P_i(\sigma)$ the set of states thought possible by girl $G_i$ when the hat distribution is
actually $\sigma$. So we have, for example,
$$
P_1(\sigma_3) =\{\sigma_3\}, P_2(\sigma_2) =\{\sigma_2\}, P_3(\sigma_4) =\{\sigma_4\}.
$$
Consequently, in each of the states $\sigma_2,\sigma_3,\sigma_4$, at least one girl would raise her hand
at the second count and conclude confidently that her hat is \emph{red}, which would signal the state (and
hence the hat distribution) to the other girls.

\medskip
If no hand goes up at the second count, \emph{all} girls know that they are in state $\sigma_1$ and will
raise their hands at the third count.

\medskip
In contrast, consider the other extreme scenario and assume:
\begin{itemize}
\item[(I')] Three girls, $G_1$, $G_2$ and $G_3$, with \emph{white} hats sit in a circle.
\item[(II)] Each girl knows that their hats are either \emph{red} or \emph{white}.
\item[(III)] Each girl can see the color of all hats except her own.
\end{itemize}

\medskip
The effect of the teacher's announcement is quite different:

\begin{itemize}
\item Each girl will immediately {conclude} that her hat is red and raise her hand because she sees only
    white hats on the other girls.
\end{itemize}

\medskip
This analysis shows:

\medskip
\begin{tcolorbox}
\begin{itemize}
\item[(i)] The information supplied by the teacher is \emph{subjective}: Even when the information
    ("there is at least one red hat") is false, the girls will eventually conclude with confidence that
    they know their hat's color.
\item[(ii)] When a girl \emph{thinks} she knows her hat's color, she may never\-theless have arrived at
    a factually wrong conclusion.
\end{itemize}
\end{tcolorbox}

\medskip
\begin{ex}\label{ex.red-hats} Assume an arbitrary distribution of red and white hats among the three
girls. Will the teacher's announcement nevertheless lead the girls to the belief that they know the color
of their hats?
\end{ex}

\subsection{Information and knowledge functions}\label{sec:information-knowledge-functions}

\subsection{Common knowledge}\label{sec:common-knowledge}

\addtocontents{toc}{\protect\newpage}
\part{n-Person Games}

\chapter{Potentials, Utilities and Equilibria}\label{chap:utilities-potentials-equilibria}

\begin{tcolorbox}
{\small Before discussing $n$-person games {\it per se}, it is useful to go back to the fundamental model
of a game $\Gamma$  being played on a system $\mathfrak S$ of states and look at characteristic features
of $\Gamma$. The aim is a general perspective on the numerical assessment of the value of states and
strategic decisions.}
\end{tcolorbox}

\section{Potentials and Utilities}\label{sec:utilities-potentials}

\subsection{Potentials}
 To have a 'potential' means to have the capability to enact
something. In physics, the term \emph{potential}\label{potential} refers to a characteristic quantity of a system whose
change results in a dynamic behavior of the system. Potential energy, for example, may allow a mass to be
set into motion. The resulting \emph{kinetic energy} corresponds to the change in the potential. Gravity
is thought to result from changes in a  corresponding potential, the so-called \emph{gravitational field},
and so on.

\medskip
Mathematically, a potential is represented as a real-valued numerical para\-meter. In other words: A
\emph{potential} on the system $\mathfrak S$ is  a function
$$
     v:\mathfrak S \to \R
$$
which assigns to a state $\sigma\in \mathfrak S$ a numerical value $v(\sigma)$. Of interest is the change in the potential resulting from a state transition $\sigma\to \tau$:
$$
\partial v(\sigma,\tau) = v(\tau) - v(\sigma)
$$

\medskip
In fact, up to a constant, the potential $v:\mathfrak S\to \R$ is determined by its \emph{marginal potential}\index{potential! marginal} $\partial v:\mathfrak S\times \mathfrak S\to \R$:

\medskip
\begin{tcolorbox}
\begin{lemma}\label{l.marginal-potential} For any potentials  $v,w:\mathfrak S\to \R$, the two statements are equivalent:
\begin{enumerate}
\item $\partial v = \partial w$.
\item There exists a constant $K_0\in \R$ such that for all $\sigma\in \mathfrak S$,
          $$w(\sigma) = v(\sigma)+K_0.$$
\end{enumerate}
\end{lemma}
\end{tcolorbox}

\Pf In the case (2), one has
$$
\partial(\sigma,\tau) = v(\tau)-v(\sigma)= w(\tau)-w(\sigma) = \partial w(\sigma,\tau)
$$
and  therefore (1). Conversely, if (1) holds, choose any $\sigma_0$ and set $K_0 = w(\sigma_0)-v(\sigma_0)$.  Then for all $\sigma\in \mathfrak S$, property (2) is apparent:
\begin{eqnarray*}
w(\sigma) &=& w(\sigma_0) +\partial w(\sigma_0,\sigma) \\
&=& K_0 +v(\sigma_0) +\partial v(\sigma_0,\sigma) \;=\; K_0 +v(\sigma).
\end{eqnarray*}

\qed

\medskip

\subsection{Utilities}
\section{Equilibria}\label{sec:equilibria}\index{equilibria}
When we talk about an ''{equilibrium}'' of an utility measure $U\in \R^{\mathfrak S\times \mathfrak S}$ on the
system $\mathfrak S$, we make the prior assumption that each state $\sigma$ has associated a
\emph{neighborhood}\index{neighborhood}
$$
   \cF^\sigma  \subseteq \mathfrak S \quad\mbox{with $\sigma \in \cF^\sigma$}
$$
and that we concentrate on state transitions to neighbors, \emph{i.e.}, to transitions of type $\sigma\to
\tau$ with $\tau\in \cF^\sigma$.

\medskip
We now say that a system state $\sigma\in \mathfrak S$ is a \emph{gain equilibrium} of $U$ if no feasible transition $\sigma\to \tau$ to a neighbor state $\tau$ has a positive utility, \emph{i.e.}, if
$$
     U(\sigma,\tau) \leq 0 \quad\mbox{holds for all $\tau\in \cF^\sigma$.}
$$
Similarly, $\sigma$ is a \emph{cost equilibrium} if
$$
     U(\sigma,\tau) \geq 0 \quad\mbox{holds for all $\tau\in \cF^\sigma$.}
$$

\medskip
\begin{remark}[Gains and costs]\label{r.gains-costs} The negative $C=-U$ of the utility measure $U$ is also a utility measure and one finds:
$$
\mbox{$\sigma$  is a gain equilibrium of $U$ }\;\Longleftrightarrow\;  \mbox{$\sigma$  is a cost
equilibrium of $C$ }
$$
From an abstract point of view, the theory of gain equilibria is equivalent to the theory of cost equilibria.
\end{remark}

\medskip
Many real-world systems appear to evolve in dynamic processes that eventually settle in an equilibrium
state (or at least approximate an equilibrium) according to some utility measure. This phenomenon is
strikingly observed in physics. But also economic theory has long suspected that economic systems may tend
towards equilibrium states\footnote{\textsc{A.A. Cournot} (1838-1877)~\cite{Cournot1838}}.

\subsection{Existence of equilibria}\label{sec:equilibria-existence}

\chapter{n-Person Games}\label{chap:n-person-games}

\begin{tcolorbox}
{\small $n$-person games generalize $2$-person games. Yet, it turns out that the special techniques for
the analysis of $2$-person games apply in this seemingly wider context as well. Traffic systems, for
example, fall into this category naturally. }
\end{tcolorbox}

\medskip
The model of a \emph{$n$-person game} $\Gamma$ assumes the presence of a finite set $N$ with $n=|N|$
elements together with a family
$$
\cX =\{X_i\mid i\in N\}
$$
of $n$ further nonempty  sets $X_i$. The elements $i\in N$ are thought of as \emph{players} (or
\emph{agents} {\it etc.}). A member $X_i\in \cX$ represents the collection of \emph{resources} (or
\emph{actions}, \emph{strategies}, \emph{decisions} {\it etc.} ) that are available to agent $i\in  N$.

\medskip
A \emph{state} of $\Gamma$ is a particular selection  $\vx = (x_i\mid i\in N) $
of individual resources $x_i\in X_i$ by the $n$ agents $i$. So the collection of all states $\vx$ of
$\Gamma$ is represented by the direct product
$$
\mathfrak X = \prod_{i\in N} X_i.
$$

\medskip
It is furthermore assumed that each player $i\in N$ has an individual \emph{utility function}\index{utility! function}
$$
 u_i  :\mathfrak X\to \R
$$
by which its individual utility of any $\vx\in \mathfrak X$ is assessed.  The whole
context
$$
      \Gamma = \Gamma(u_i\mid i\in N )
$$
now describes the $n$-person game under consideration.

\medskip
\begin{ex}\label{ex.knobeln} The matrix game $\Gamma$ with a row player $R$ and a column player $C$ and
the payoff matrix
$$
    P =\begin{bmatrix} (p_{11},q_{11})  &(p_{12},q_{12}))\\ (p_{21},q_{21}) &(p_{22},q_{22}) \end{bmatrix}
    = \begin{bmatrix} (+1,-1) &(-1,+1)\\ (-1,+1) &(+1,-1)\end{bmatrix}.
$$
is a $2$-person game with the player set $N= \{R,C\}$ and the strategy sets $X_R = \{1,2\}$ and $X_C =
\{1,2\}$. Accordingly, the set of states is
$$
\mathfrak X = X_R\times X_C = \{(1,1), (1,2),(2,1),(2,2)\}.
$$
The individual utility functions $u_R, u_C:\mathfrak X\to \R$  take the values
$$
   u_R(s,t) = p_{st} \quad\mbox{and} \quad  u_C(s,t) = q_{st} \quad \mbox{for all
   $(s,t)\in \mathfrak X$.}
$$
\end{ex}

\medskip
\begin{remark} It is often  convenient  to label  the elements of $N$ by natural numbers and assume
$N=\{1,2,\ldots,n\}$ for simplicity of notation. In this case, a state $\vx$ of $\Gamma$ can be denoted in
the form
$$
\vx = (x_1,x_2,\ldots,x_n) \in X_1\times X_2\times \ldots\times X_n.
$$
\end{remark}

\medskip
\paragraph{\bf Cooperation.}\index{cooperation} The basic game model with a set $N$ of players is readily
generalized to a model where groups of players (and not just individuals) derive a utility value from a
certain state $\vx\in \mathfrak X$.  To this end, we call a subset $S\subseteq N$ of players a
\emph{coalition}\index{coalition} and assume an individual utility function $u_S:\mathfrak X\to \R$ to
exist for each coalition $S$.

\medskip
From an abstract mathematical point of view, however, this generalized model can be treated like a
standard $|\cN|$-person game, having the set
$$
      \cN = \{S\subseteq N\}
$$
of coalitions as its set of ''superplayers''. In fact, we may allow each coalition $S$ to be endowed with
its own set $X_S$ of resources. In this chapter, we therefore retain the basic model with respect to an
underlying  set $N$ of players.

\medskip
Further aspects come to the fore, however, when one asks what the strategic decisions at coalition level
mean for the individual players. For example:
\begin{itemize}
\item \emph{How should one assess the power of an individual player?}
\item \emph{How do coalitions come about?}
\end{itemize}

A special class of potential $n$-person games with cooperation, so-called
\emph{TU-games}, will be studied in their own right in more detail in
Chapter~\ref{chap:Cooperative-games}.

\medskip
\paragraph{\bf Probabilistic models.}  There are many probabilistic aspects of $n$-person games. One
consists in having a probabilistic model for the choice of actions to start with  (see
Ex.\ref{ex.fuzzy-games}).

\medskip
\begin{ex}[Fuzzy games]\label{ex.fuzzy-games}\index{game!fuzzy} Assume a game $\Gamma$ where any player
$i\in N$ has to decide between two alternatives, say  "0" and "1",  and chooses "1" with probability
$x_i$. Then $\Gamma$ is a $|N|$-person game in which each player $i$  has the unit interval
$$
     X_i =[0,1] = \{x\in \R\mid 0\leq x\leq 1\}
$$
as its set of resources. A  joint strategic choice
$$\vx=(x_1,\ldots, x_i,\ldots,x_n)\in [0,1]^N
$$
can be interpreted as a ''fuzzy'' decision to form a coalition  $X\subseteq N$:
\begin{itemize}
\item Player $i$  will be a member of $X$ with probability $x_i$.
\end{itemize}
$\vx$ is thus  the description of a \emph{fuzzy coalition}\index{coalition!fuzzy}\index{fuzzy}. $\Gamma$
is a \emph{fuzzy cooperative game} in the sense of \textsc{Aubin}~\cite{Aubin1981}.
\end{ex}

\medskip
A further model  arises from  the \emph{randomization} of a $n$-person game (see
Section~3 below).  Other probabilistic aspects of $n$-person games are studied in
Chapter~\ref{chap:Cooperative-games} and in Chapter~\ref{chap:Interaction-quantum-models}.

\section{Dynamics of $n$-person games}\label{sec:n-person-dynamics}

\section{Equilibria}\label{sec:n-equilibria}

\section{Randomization of matrix games}\label{sec:n-matrix-games}

\section{Traffic flows}\label{sec:traffic-flows}\index{\textsc{Wardrop}}\index{traffic!flow}

\chapter{Potentials and Temperature}\label{chap.Potentials-Temperature}

\begin{tcolorbox}
{\small The temperature of a system depends on the measuring device in use, which is mathematically represented
as a potential function. \textsc{Boltzmann}'s approach to the notion of temperature in statistical
thermodynamics extends to general systems. Of particular interest are $n$-person matrix games where the
temperature reflects the activity of the player set as a whole with respect to the total utility. The
interpretation of the activity as a \textsc{Metropolis} process moreover indicates how the strategic
decisions of individual players influence the expected value of the measuring device.}
\end{tcolorbox}

\medskip
Consider a finite system $\mathfrak S$ that is in a state $\sigma$ with probability $\pi_\sigma$. Then
$\mathfrak S$ has the entropy
$$
   H(\pi) = \sum_{\sigma\in \Sigma} \pi_\sigma\ln(1/\pi_\sigma) = -\sum_{\sigma\in \Sigma} \pi_\sigma\ln
   \pi_\sigma.
$$

The expected value of a potential $v\in \R^{\mathfrak S}$ will be
$$
     E(v,\pi) = \sum_{\sigma\in \mathfrak S} v_\sigma\pi_\sigma.
$$

\medskip
Let us think of $v$ as a numerical measuring device for a certain characteristic feature of $\mathfrak S$.
In a physical model, the number $v_\sigma$ could describe the level of inherent ''energy'' of $\mathfrak S$ in the state $\sigma$, for example.  In economics, the function $v:\mathfrak S\to \R$ could be a representative statistic for the general state of the economy. In the context of a $n$-person game, $v_\sigma$ could possibly measure a degree of ''activity'' of the set $N$ of players in the state $\sigma$ {\it etc.}

\medskip
Of course, the activity level $v_\sigma$ depends on the particular characteristic feature that is
investigated under $v$. Different features of $\mathfrak S$ may display different activity levels in the
same state $\sigma$.

\section{Temperature}\label{sec:temperature}

\subsection{\textsc{Boltzmann} temperature}\label{sec:Boltzmann-temperature}

From Lemma~\ref{l.Boltzmann-expectation}, it is clear that one could characterize the expected value $\mu$
of a non-constant potential $v$ equally well by specifying the parameter $t\in \R\cup\{-\infty,+\infty\}$ of the
\textsc{Boltzmann} distribution $\beta(t)$ with expectation
$$
      \mu(t) = \mu.
$$

In analogy with the \textsc{Boltzmann} model in statistical thermodynamics for the temperature, we call
the related parameter
$$
      T = 1/t.
$$
the \emph{temperature}\index{temperature} of the system $\mathfrak S$ relative to a potential with the
expected value $\mu(1/T)$. Adjusting the notation accordingly to
$$
   \beta^{(T)} = \beta(1/T) \quad\mbox{and}\quad \mu^{(T)} = \mu(1/T).
$$

the \textsc{Boltzmann} distribution $\beta^{(T)}$ has the coefficients
$$
  \beta^{(T)}_\sigma = \frac{e^{v_\sigma/T}}{\sum_{\tau\in \mathfrak S} e^{v_\tau/T}} \quad(\sigma\in
  \mathfrak S).
$$

\medskip
As the system ''freezes'' to the temperature $T=0$, one obtains the extreme values of the potential $v$ as the
expectations in the limit, depending on whether the limit $0$ is approached from the positive or the
negative side:

\begin{eqnarray*}
\D\lim_{T\to 0^+} \mu^{(T)} &=& \max_{\sigma\in \mathfrak S} v_\sigma\\
\D\lim_{T\to 0^-} \mu^{(T)} &=& \min_{\sigma\in \mathfrak S} v_\sigma.
\end{eqnarray*}

\medskip
In contrast, all states of $\mathfrak S$ are equally likely at when the temperature $T$ is infinite.

\section{The \textsc{Metropolis} process}\label{sec:Metropolis-process}

\section{Temperature of matrix games}\label{sec:temperature-matrix-games}
Let $\Gamma=\Gamma(u_i\mid i\in N)$ be a $n$-person game with player set $N=\{1,\ldots,n\}$ where each
player $i\in N$ has a finite set $X_i$ of strategic resources and an utility function
$$
u_i:\mathfrak X\to \R \quad\mbox{(with $\mathfrak X = X_1\times X_2\times \ldots\times X_n$).}
$$

\medskip
In the model of randomized matrix games, it is assumed that the players $i$ choose probability
distributions $\pi^{(i)}$ on their strategy sets $X_i$ \emph{independently from each other} and then
select elements $x_i\in X_i$ according  to those distributions.

\medskip
Let us drop the stochastic independence assumption and consider the more general model where the joint
strategy
$$
    \vx =(x_1,x_2,\ldots,x_n) \in \mathfrak X
$$
would be chosen by the player set $N$ with a certain probability $\pi_\vx$. The aggregated total utility
value is then expected to be
$$
\mu = \sum_{\vx\in \mathfrak X}\sum_{i\in N} u_i(\vx) \pi_\vx.
$$

The players' total utility
$$
   u(\vx) = \sum_{i\in N} u_i(\vx)
$$
is a potential on $\mathfrak X$. So one may consider the (\textsc{Boltzmann}) temperature relative to $u$. In the case
$$
   \mu = \frac{1}{Z_T} \sum_{\vx\in \mathfrak X} e^{u(\vx)/T} \quad(\mbox{with $Z_T = Z(1/T)$})
$$
we say that $\Gamma$ is \emph{is played at temperature $T$}. If $|T|\approx \infty$ (\emph{i.e.}, $|T|$ is
very large), we expect about the average value of the total utility:
$$
   \mu \approx \frac{1}{|\mathfrak X|}\sum_{\vx\in \mathfrak X} u(\vx).
$$

\medskip
If $T>0$ is very small (\emph{i.e.}, $T\approx 0$), then we may expect about the maximal total utility:
$$
  \mu \approx \max_{\vx\in \mathfrak X}  u(\vx).
$$

\medskip
Similarly, if $T\approx 0$ and $T< 0$ holds, about the minimal total utility value is to be expected:
$$
  \mu \approx \min_{\vx\in \mathfrak X}  u(\vx).
$$

\medskip
\begin{tcolorbox} \index{social justice}
\begin{remark}[Social justice]\label{r.social-justice} It appears to be in the joint interest of the
community $N$ of players to play $\Gamma$ at a temperature $T$ that is close to $0$ but positive if a
large total utility value is desired and negative if a minimal value is sought.

\medskip
The potential function $u$ is equivalent (up to the scaling factor $n$) to the average utility function $\ov{u}$ of the members of~$N$:
$$
   u(\vx) = \sum_{i=1}^n u_i(\vx) \quad \longleftrightarrow \quad \ov{u}(\vx) =   \frac1n\sum_{i=1}^n
   u_i(\vx).
$$

A high group average does not necessarily imply a guaranteed high utility value for \emph{each} individual
member in $N$, however.

\medskip
{\rm To formulate it bluntly:}
\begin{itemize}
\item Even when  a high average utility value is used as a criterion for ''social justice'' in $N$,
    there may still be members of $N$ that are not treated ''fairly''.
\end{itemize}
\end{remark}
\end{tcolorbox}

\medskip
The interplay of different interests (individual utility  of the players vs. combined utility of the set
of all players) is studied in more detail within the framework of \emph{cooperative games} in
Chapter~\ref{chap:Cooperative-games}.

\chapter{Cooperative Games}\label{chap:Cooperative-games}

\begin{tcolorbox}
{\small Players in a cooperative game strive for a common goal, from which they possibly profit. Of special interest is the class of TU-games with a transferable utility potential, which is best studied within the context of linear algebra. Central is the question how to distribute the achieved goal's profit appropriately. The core of a
cooperative game is an important analytical notion. It strengthens the \textsc{von
Neumann-Morgenstern} solution concept of stable sets and provides a link to the theory of discrete
optimization and greedy algorithms. It turns out that the core is the only stable set in so-called
supermodular games. Values of cooperative games are more general solution concepts and can be motivated by stochastic models for the formation of coalitions.
Natural models for the dynamics of coalition formation are closely related to thermodynamical models in
statistical physics and offer an alternative view on the role of equilibria.}
\end{tcolorbox}

\medskip
While the agents in the $n$-person games of the previous chapters typically have individual utility
objectives and thus possibly opposing strategic goals, the model of a \emph{cooperative
game}\index{game!cooperative} refers to a finite set $N$ of $n=|N|$ players that may or may not be active
towards a common goal.  A subset $S\subseteq N$ of potentially active players is traditionally called a
\emph{coalition}\index{coalition}. Mathematically, there are several ways of looking at the system of
coalitions:

\medskip
From a set-theoretic point of view, one has the system of the $2^n$ coalitions
$$
\cN= \{S\mid S\subseteq N\}.
$$
On the other hand, one may represent a subset $S\in \cN$ by its incidence vector $x^{(S)}\in \R^N$ with
the coordinates
$$
      x^{(S)}_i  = \left\{\begin{array}{cl} 1 &\mbox{if $i\in S$}\\ 0 &\mbox{if $i\notin
      S$.}\end{array}\right.
$$
The incidence vector $x^{(S)}$ suggests the interpretation of an ''activity vector'':
$$
\mbox{$i\in N$ is \emph{active} if $x^{(S)}_i=1$.}
$$
The coalition $S$ would thus be the collection of active players.

\medskip
A further interpretation imagines every player $i\in N$ to have a binary strategy set $X_i=\{0,1\}$ from
which to choose one element. An incidence vector
$$
     x = (x_1,\ldots,x_n)\in X_1\times \cdots\times X_n = \{0,1\}^N \subseteq \R^N
$$
represents the joint strategy decision of the $n$ players and we have the correspondence
$$
    \cN \quad \longleftrightarrow \quad \{0,1\}^N = 2^N
$$

\medskip
By a \emph{cooperative game} we will just understand a $n$-person game $\Gamma$ with player set $N$ and
state set
$$
     \mathfrak X = \cN \quad\mbox{or}\quad \mathfrak X = 2^N,
$$
depending on a set-theoretic or on a vector space point of view. A general cooperative game $\Gamma=(u_i\mid i\in N)$ with individual utility functions $u_i: \cN\to \R$ is therefore a matrix game where each player has the choice between two alternative actions.

\medskip
In the following, we will concentrate on cooperative games whose individual utilities are implied  by a general potential on $\cN$.

\section{Cooperative TU-games}\label{sec:cooperative-TU-games}\index{TU-game}

\section{Vector spaces of TU-games}\label{sec:spaces-TU-games}

\subsection{\textsc{M\"obius} transform}\label{sec:Moebius-transformation}

\subsection{\bf Potentials and linear functionals}\label{sec:potential-linear-functional} A potential
$f:\cN\to \R$, inter\-preted as a vector $f\in \R^\cN$ defines a linear functional
$\tilde{f}:\R^\cN\to\R$ with the values
$$
\tilde{f}(g) = \(f|g\) = \sum_{S\in \cN} f_Sg_S\quad \mbox{for all $g\in \R^\cN$.}
$$
If $g^{(S)}$ is the $(0,1)$-incidence vector of a particular coalition $S\in \cN$, we have
$$
  \tilde{f}(g^{(S)}) =\(f|g^{(S)}\)   = f_S\cdot 1 = f_S,
$$
which means that $\tilde{f}$ extends the potential $f$ on $2^N$ ($=\cN$) to all of $\R^\cN$.

\medskip
Conversely, every linear functional $g\mapsto \(f|g\)$ on $\R^\cN$ defines a unique potential $f$ on $\cN$
{\it via}
$$
    f(S) = \(f|g^{(S)}\) \quad\mbox{for all $S\in \cN$.}
$$

These considerations reveal characteristic functions on $\cN$ and linear functionals on
$\R^\cN$ to be two sides of the same coin. From the point of view of linear algebra, one can therefore
equivalently define:

\medskip
\begin{tcolorbox}
\begin{itemize}
\item \emph{A cooperative TU-game is a pair $\Gamma=(N,v)$, where $N$ is a set of players and $v\mapsto
    \(v|g\)$ is a linear functional on the vector space $\R^\cN$.}
\end{itemize}
\end{tcolorbox}

\subsection{Marginal values}\label{sec:marginal-values} 

\section{Examples of TU-games}\label{sec:examples-TU-games}

\subsection{Additive games}\label{sec:additive-games}\index{game!additive} The marginal value

\subsection{Production games}\label{sec:production-games}\index{game!production} Similar to the situation in Chapter~\ref{sec:shadow-price}, consider a set $N$ of players in an economic production environment where  there are $m$ raw materials, $M_1,\ldots, M_m$ from which goods of $k$ different types may be manufactured.

\medskip
Let $x=(x_1,\ldots,x_k)$ be a plan that proposes the production of $x_j\geq 0$ units of the $j$th good and assume:
\begin{enumerate}
\item $x$ would need $a_i(x)$ units of material $M_i$ for all $i=1,\ldots,m$;
\item each supplier $s\in N$ has $b_{is}\geq 0$ units of material $M_i$ at its disposal;
\item the production $x$  could be sold for the price of $f(x)$.
\end{enumerate}

\medskip
So the coalition $S\subseteq N$ could guarantee a production of market value
\begin{equation}\label{eq.production-game}
v(S) = \max_{x\in \R^k_+}~ f(x) \quad\mbox{s.t.}\quad a_i(x)\leq \sum_{s\in S} b_{is} \;(i=1,\ldots,m).
\end{equation}

\medskip
The corresponding cooperative TU-game $(N,v)$ is a \emph{production game}.

\medskip
\paragraph{\bf What is the worth of a player?} This is one of the central questions in cooperative game theory. In the context of the production game $(N,v)$, one natural approach to resolve this question is the market price principle:

\medskip
\begin{tcolorbox}
\begin{itemize}
\item[(MP)] Assuming that each material $M_i$ has a market price of $y_i$ per unit, assign to each supplier $s\in N$ the market value $w_s$ of its inventory:
$$
w_s = \sum_{i=1}^m y_i b_{is} .
$$
\end{itemize}
\end{tcolorbox}

\medskip
An objection against a simple application of the principle (MP) could possibly be made if
$$
     v(S) > \sum_{s\in S} w_s \quad\mbox{holds for some coalition $S\subseteq N$.}
$$
In this case, $S$ could generate a market value that is strictly larger than the market value of its inventory. So the intrinsic economic value of the members of $S$ is actually larger than the value of their inventory. This consideration leads to another worth assessment principle:

\medskip
\begin{tcolorbox}
\begin{itemize}
\item[(CA)] Assign numbers $w_s$ to the members of $N$ such that
$$
v(N) = \sum_{s\in N} w_s \quad\mbox{and}\quad v(S) \leq \sum_{s\in S} w_s \;\mbox{for all $S\subseteq N$}.
$$
\end{itemize}
\end{tcolorbox}

\medskip
An allocation $w\in \R^N$ according to principle (CA) is a so-called \emph{core allocation}. Core allocations do not necessarily exist in a given cooperative game, however.\footnote{core allocations  are studied more generally in Section~5}

\medskip
As it turns out, the principles (MP) and (CA) can be satisfied simultaneously if the production game $(N,v)$ has a linear objective and linear restrictions.

\medskip
\paragraph{\bf Linear production games.} Assume that the production game with characteristic function (\ref{eq.production-game}) is \emph{linear} in the sense
$$
\begin{array}{cccclc}
f(x) &=& c^Tx &=& c_1x_1 +\ldots + c_nx_k \\
a_i(x)&=& a_i^Tx  &=& a_{i1}x_1 + \ldots + a_{in} x_k &(i=1,\ldots,m)
\end{array}
$$
and admits an optimal production plan $x^*$ with market value
$$
v(N)= f(x^*) = c^Tx^*.
$$
$x^*$ is the solution of a linear program. So also an optimal solution $y^*$ exists for the dual linear program
$$
    \min_{y\in \R^m_+} \sum_{i=1}^m b^N_iy_i  \quad\mbox{s.t.}\quad  \sum_{i=1}^m a_{ij}y_i  \geq c_j \quad(j=1,\ldots,k),
$$
where we have used the notation for the aggregated inventory of the members of a coalition:
$$
     b^S_i = \sum_{s\in S} b_{is}  \quad\mbox{for any $S\subseteq N$}.
$$
The components $y^*_i$ of $y^*$ are the shadow prices of the materials $M_i$. According to principle (MP), let us allocate the individual worth
$$
    w_s^* =\sum_{i=1}^m y^*_i b_{is} \quad\mbox{to any $s\in N$.}
$$

To see that $w^*$ satisfies also the principle (CA), observe first from linear programming duality:
$$
 \sum_{s\in N} w_s^* =  \sum_{i=1}^m b^N_iy_i^* = \sum_{j=1}^k c_jx^*_j  = v(N).
$$

The dual of any $S$-restricted production problem (\ref{eq.production-game}) differs only in the coefficients of the objective function. Expressed in terms of the dual linear program, one has
$$
 v(S) = \min_{y\in \R^m_+} \sum_{i=1}^m b^S_i y_i \quad\mbox{s.t.}\quad \sum_{i=1}^m a_{ij}y_i  \geq c_j \quad(j=1,\ldots,k).
$$
Since $y^*$ is a feasible (although not necessarily optimal) dual solution for any $S$-restricted problem, one concludes:
$$
 v(S) \;\leq\; \sum_{i=1}^m b^S_iy_i^* = \sum_{s\in S}^m w_s^*.
$$

\subsection{Network connection games}\label{sec:network-games}\index{game!network}

\subsection{Voting games}\label{sec:voting-games} \index{game!threshold}\index{game! voting}\index{voting game}
Assume there is a set $N$ of $n$ voters $i$ of not necessarily equal voting power. Denote by $w_i$ the
number of votes voter $i$ can cast. Given a threshold $w$, the associated \emph{voting game}\footnote{also
known as a \emph{threshold game}}  has the characteristic function
$$
   v(S) = \left\{\begin{array}{cl} 1 &\mbox{if $\D\sum_{i\in S} w_i \geq w$} \\ 0
   &\mbox{otherwise.}\end{array}\right.
$$

In the voting context, $v(S)=1$ has the interpretation that the coalition $S$ has the voting power to make
a certain proposed measure pass. Notice that in the case $v(S) = 0$, a voter $i$ with marginal value
$$
      \partial_iv(S) = v(S\cup i)-v(S) = 1
$$
has the power to swing the vote by joining $S$. The general question is of high political importance:

\medskip
\begin{tcolorbox}
\begin{itemize}
\item \emph{How can (or should) one assess the overall voting power of a voter $i$ in a voting context?}
\end{itemize}
\end{tcolorbox}

\medskip
\begin{remark}\label{r.Banzhaf}
A popular index for individual voting power is the \textsc{Banzhaf} power index (see
Section~8 below). However, there are alternative evaluations that also have their
merits. As in the case of network cost allocation, abstract mathematics cannot decide what the ''best''
method would be.
\end{remark}

\section{Generalized coalitions and balanced games}\label{sec:generalized-coalitions-balanced games}
Let us assume that the TU-game $(N,v)$ can be played by several coalitions $S\subseteq N$
''simultaneously", requiring an activity level $y_S\geq 0$ from every member $i\in S$ so that no player has
to invest more than $100\%$ of its available activity resources in total. With this in mind, we define a \emph{generalized
coalition}\footnote{also known as a \emph{packing}} \index{packing}  to be a nonnegative vector \index{coalition! generalized}
$$
   \vy = (y_S\mid S\subseteq N) \in \R^\cN_+ \quad\mbox{s.t.}\quad \sum_{S\ni i} y_S \leq 1  \;\forall
   i\in N
$$
and associate with it the utility value
$$
    v(\vy) = \(v|\vy\) = \sum_{S\subseteq N} v(S)y_S.
$$

\medskip
\begin{ex}\label{ex.disjoint-sets} Assume that $\vy =(y_S|S\subseteq N)$ is a generalized coalition with
binary components $y_S\in \{0,1\}$. Show that $\vy$ is the incidence vector of a family of pairwise
disjoint coalitions.
\end{ex}

\medskip
\begin{ex}[Fuzzy coalitions]\label{ex.fuzzy-coalitions}\index{coalition! fuzzy} Let $\pi=(\pi_S| S\subseteq N)$ be a probability dis\-tribution on the family $\cN$ of all coalitions. Then one has $\pi_S\geq 0$ for all $S\in \cN$ and
$$
    \sum_{S\ni i} \pi_S \leq \sum_{S\in \cN} \pi_S = 1 \quad\mbox{for all $ i\in N.$}
$$
So $\pi$ represents a generalized coalition which generalizes the notion of a fuzzy coalition in the sense of Ex.~\ref{ex.fuzzy-games}.
\end{ex}

\medskip
Denote by $\cY\subseteq \R^\cN_+$ the collection of all generalized coalitions $\vy$ and note that $\cY$
is a non-empty, convex and compact set. The optimal utility value $\ov{v}$ is the optimal solution of a
feasible linear program:
\begin{equation}\label{eq.GenLP1}
\ov{v} = \max_{\vy\in \cY} v(\vy) = \max_{\vy\in \R^\cN_+}  \sum_{S\subseteq N} v(S)y_S \quad\mbox{s.t.}
\sum_{S\ni i}y_S\leq 1 \;\forall i\in N.
\end{equation}

Taking $\vy^N$ as the vector with components $y_N = 1$ and $y_S = 0$ if $S\neq N$, we see immediately:
$$
     \ov{v}\geq v(\vy^N) = v(N).
$$
The game $(N,v)$ is called \emph{(positively) balanced}\index{balanced game}\index{game! balanced} if
equality is achieved:
$$
     \ov{v} = v(N).
$$

\medskip
The dual linear program associated with (\ref{eq.GenLP1}) has the same optimal value:
\begin{equation}\label{eq.GenLP2}
\ov{v} = \min_{x\in \R^N_+} \sum_{i\in N} x_i \quad\mbox{s.t.} \sum_{i\in S}x_i\geq v(S) \;\forall S
\subseteq  N.
\end{equation}

\medskip
Hence linear programming\footnote{\emph{cf.} Theorem~\ref{t.main-LP}} duality yields:

\medskip
\begin{tcolorbox}
\begin{theorem}[\textsc{Bondareva}~\cite{Bondareva1963}]\label{t.Bondareva} For any cooperative game
$(N,v)$, the two statements are equivalent:
\begin{enumerate}
\item $(N,v)$ is (positively) balanced.\index{balanced! positively}
\item  For each $i\in N$ there is a number $x_i\geq 0$ such that
$$
v(N) = \sum_{i\in N}x_i \quad\mbox{and}\quad    \sum_{i\in S} x_i \geq v(S) \quad\mbox{for all
$S\subseteq N$.}
$$
\end{enumerate}
\end{theorem}
\end{tcolorbox}

\medskip
\begin{ex}\label{ex.balanced-grand-coalition} Let $(N,v)$ be a balanced game. Show:
$$
v(N) = \max_{S\subseteq N} v(S).
$$
\end{ex}

\medskip
\begin{ex}[Linear production games]\label{ex.balanced-LP-games} Show that a linear production game is positively balanced if and only if it admits an optimal production plan.
\end{ex}

\medskip
\paragraph{\bf Covers.}\index{cover} The generalized coalition $\vy=(y_S|S\subseteq N)$ is said to \emph{cover} the set $N$ if equality
$$
   \sum_{S\ni i} y_S = 1 \quad\mbox{holds for all elements $i\in N$,}
$$
which means that each agent $i$'s activity resource of unit value $1$ is fully used under $\vy$. The \emph{covering value}\index{covering value}\index{value! covering} of $(N,v)$ is the number
$$
   v^c = \max~\{v(\vy) \mid \mbox{$\vy$ is a cover of  $(N,v)$}\}.
$$

As in the derivation of Theorem~\ref{t.Bondareva}, we can characterize the covering value by linear programming duality and find

\medskip
\begin{tcolorbox}
\begin{equation}\label{eq.covering-value}
v^c = \min_{x\in \R^N} \sum_{i\in N} x_i\quad\mbox{s.t.}\quad \sum_{i\in S}x_i\geq v(S) \;\forall S\subseteq N.
\end{equation}
\end{tcolorbox}

\medskip
\begin{ex}\label{ex.strongly-balanced} Prove formula (\ref{eq.covering-value}).
\end{ex}

\medskip
Clearly, one has $v(N) \leq v^c \leq \ov{v}$. Calling the game $(N,v)$ \emph{strongly balanced}\index{balanced!strongly} if it yields the equality
$$
     v(N) = v^c ,
$$
we therefore find:

\medskip
\begin{tcolorbox}
\begin{proposition}\label{p.Bondareva} Every positively balanced game is strongly balanced.
\end{proposition}
\end{tcolorbox}

\section{The core}\label{sec:core}

\subsection{Stable sets}\label{sec:stable-sets}

\subsection{The core}\label{sec:the-core}
Say that the payoff $x\in \R^N$ is \emph{coalition rational}\index{coalition!rational} in the game $(N,v)$
if each coalition is awarded at least its own value, \emph{i.e.}, if
$$
    x(S) \geq v(S) \quad\mbox{holds for all $S\subseteq N$.}
$$

\medskip
The \emph{core}\index{core} of a cooperative profit game $ (N,v)$  is the set of all feasible coalition
rational payoff vectors:
$$
   \core(v) = \{x\in \R^N \mid x( N) \leq  v(N), x(S) \geq v(S) \;\forall S\subseteq N \}.
$$

\medskip
\begin{remark}[Efficiency]\label{r.efficient} Note that every payoff vector $x\in \core(v)$ is
\emph{efficient}\index{efficient} in the sense
$$
     x(N) = v(N).
$$
\end{remark}

\medskip
\begin{ex}\label{ex.core-domination} Let $x,y\in \core(v)$. Then $x$ cannot dominate $y$ because otherwise
a coalition $S$ would exist with the property
$$
v(S) \geq x(S) > y(S) \geq v(S),
$$
which is a mathematical contradiction.
\end{ex}

\medskip
\begin{tcolorbox}
\begin{proposition}\label{p.core-stable} Let $\cS$ be an arbitrary stable set of the cooperative game
$(N,v)$. Then
$$
    \core(v) \subseteq \cS.
$$
\end{proposition}
\end{tcolorbox}

\Pf Suppose to the contrary, that a vector $y\in \core(v)\setminus \cS$ exists. Since $\cS$ is stable, it contains a payoff  $x\in \cS$ that dominates $y$, \emph{i.e}, there exists a coalition $S\subseteq N$ so that
$$
v(S) \leq y(S) < x(S) \leq v(S),
$$
which is impossible.

\qed

\medskip
The core of a cost game $(N,c)$ is defined analogously:
$$
\core^*(c) = \{x\in \R^N \mid x( N) \geq c(N), x(S) \leq c(S) \;\forall S\subseteq N \}.
$$

Every allocation $x\in \core^*(c)$  distributes the cost $c(N)$ among the players $i\in N$ so that no coalition $S$ pays
more than its proper cost $c(S)$.

\medskip
\begin{ex}\label{ex.profit-core} Show for the (zero-normalized) cooperative game $(N,v)$ and its dual
$(N,v^*)$:
$$
      \core(v^*) =  \core^*(v).
$$
\end{ex}

\medskip
\begin{ex}\label{ex.no-core} Give an example of a  game $(N,v)$ with $\core(v) = \emptyset$.
\end{ex}

\medskip
\begin{tcolorbox}
\begin{proposition}\label{p.balanced-games} Let $(N,v)$ be an arbitrary TU-game. Then:
\begin{enumerate}
\item $\core(v)\neq \emptyset \quad \Longleftrightarrow \quad\mbox{$(N,v)$ is strongly balanced.}$
\item If $v(\{i\})\geq 0$ holds for all $i\in N$ in the game $(N,v)$, then
$$
    \core(v)\neq \emptyset \quad \Longleftrightarrow \quad\mbox{$(N,v)$ is positively balanced.}
$$
\end{enumerate}
\end{proposition}
\end{tcolorbox}

\Pf Exercise left to the reader (\emph{cf.} Theorem~\ref{t.Bondareva} and Proposition~\ref{p.Bondareva}).

\section{Core relaxations}\label{sec:core-relaxations}

\subsection{Nuclea}\label{sec:nuclea} The idea of the least core is a relaxation of the constraint $x(N)=v(N)$ while retaining the other core constraints $x(S)\geq v(S)$.

\medskip
An alternative approach to a relaxation of the core concept consists in retaining the equality $x(N)=v(S)$ while possibly relaxing the other constraints.

\medskip
To make the idea precise, say that $f\in \R^\cN$ is  a \emph{relaxation vector} vector if
$$
 f_\emptyset = 0 = f_N  \quad\mbox{and}\quad f_S \geq 0 \;\;\mbox{for all coalitions $S\in \cN$}.
$$
$f$ is \emph{feasible} for $v$ if there exists some scalar $\epsilon\in \R$ such that
$$
   C(f,\epsilon) = \core(v-\epsilon f) \neq \emptyset.
$$
Hence, if $\core(v)\neq \emptyset$, every relaxation vector $f$ is feasible (with $\epsilon =0$, for example).

\medskip
\begin{tcolorbox}
\begin{lemma}\label{l.f-feasible0} Assume that $f$ is a feasible relaxation with $f_S >0$ for at least one $S\in \cN$. Then there exists a scalar $\epsilon_0\in \R$ such that
$$
    C(f,\epsilon)\neq 0 \quad \Longleftrightarrow \quad \epsilon \geq \epsilon_0.
$$
\end{lemma}
\end{tcolorbox}

\subsection{Nucleolus and nucleon}\label{sec:nucleolus-nucleon} The \emph{nucleolus} \index{nucleolus} of the game $(N,v)$ introduced by
\textsc{Schmeidler}~\cite{Schmeidler1969} is the $f^1$-nucleon relative to the relaxation vector $f^1$ with the unit parameters
$$
    f^1_S = 1 \quad\mbox{for all coalitions $S\neq \emptyset, N$.}
$$

By Proposition~\ref{p.unique-nucleon}, it is clear that the nucleolus always exists and is a singleton.\footnote{related  solution concepts are studied in \textsc{Maschler} {\it et al.}~\cite{Maschler-et-al1979}}

\medskip
The \emph{nucleon}\footnote{see \textsc{Faigle} {\it et al.}~\cite{Faigle-et-al1998}} of a game $(N,v)$ with a nonnegative characteristic function is the $f^v$-nucleon of the game relative to the relaxations
$$
    x(S) \geq (1-\epsilon)v(S)  \quad\mbox{for $0\leq \epsilon \leq 1$}
$$
\emph{i.e.}, the relaxation with the coefficients $f^v(S) = v(S)$ for $S\neq N$.

\medskip
The choice $\epsilon=1$ shows that the nucleon relaxation is feasible. The nucleon is a singleton vector $x^v$ if the (incidence vectors of the) coalitions $S\in \cN$ with value $v(S)>0$ yield a system of  full rank $n$.

\subsection{Excess minimization}\label{sec:excess-minimization}

\section{\textsc{Monge} vectors and supermodularity}\label{sec:Monge-vectors}

\subsection{\bf The \textsc{Monge}  extension}\label{sec:Monge-extension}

\subsection{Linear programming aspects}\label{sec:LP-core} Generalizing the approach to the notion of balancedness of Theorem~\ref{t.Bondareva}, let us consider the linear program

\begin{equation}\label{eq.core-LP1}
\min_{x\in \R^N} c^Tx \;\;\mbox{s.t.}\;\;\mbox{$x(N) =v(N),\; x(S)\geq v(S)$ if $S\neq N$.}
\end{equation}

and its dual

\begin{equation}\label{eq.core-LP2}
\max_{y\in \R^\cN} v^Ty \;\;\mbox{s.t.}\;\;\mbox{$\D\sum_{S\ni i} y_S \leq c_i \;\forall i\in N,\;
y_S\geq 0$ if $S\neq N$.}
\end{equation}

for a given parameter vector $c\in \R^N$. Observe in the case
$$
c_{i_1}\geq \ldots \ldots \geq c_{i_n}
$$
that the dual \textsc{Monge}  vector $y^\pi$
relative to $c$ is a dually feasible solution since $y^\pi_S\geq 0$  holds for all $S\neq N$.  The feasible primal solutions, on the other hand, are exactly the members of $\core(v)$.

\medskip
Hence, if $\core(v)\neq \emptyset$, both linear programs have optimal solutions. Linear programming duality then furthermore shows

\begin{equation}\label{eq.primal-core-function}
   \tilde{v}(c) = \min_{x\in \core(v)} c^Tx \; \geq\;  v^Ty^\pi = [v](c).
\end{equation}

\medskip
\begin{tcolorbox}
\begin{theorem}\label{t.Monge-LP}   $\tilde{v} = [v]$ holds for the game $(N,v)$ if and only if all primal
\textsc{Monge}  $x^\pi$ vectors lie in $\core(v)$.
\end{theorem}
\end{tcolorbox}

\medskip
\Pf Assume $c_{i_1}\geq \ldots\geq c_{i_n}$ and $\pi = i_1\ldots i_n$. If $x^\pi\in \core(v)$, then
$x^\pi$ is a feasible solution for the linear program (\ref{eq.core-LP1}).  Since the dual \textsc{Monge}
vector $y^\pi$  is feasible for (\ref{eq.core-LP2}), we find
$$
   c^Tx^\pi \geq \tilde{v}(c) \geq [v](v) = c^Tx^\pi \quad\mbox{and hence}\quad \tilde{v}(c) = [v](c).
$$
Conversely, $\tilde{v}=[v]$ means that the dual \textsc{Monge}  vector is guaranteed to yield an optimal
solution for (\ref{eq.core-LP2}). So consider an arrangement $\psi = j_1\ldots j_n$ of $N$ and the
parameter vector $c\in \R^N$ with the components
$$
     c_{j_k}= n+1-k \quad\mbox{for $k = 1,\ldots,n$.}
$$
The dual vector $y^\psi$ has strictly positive components $y^\psi_{S_k} = 1 >0$  on the sets $S_k^\psi$. It
follows from the KKT-conditions for optimal solutions  that an optimal solution $x^*\in\core(v)$ of the
corresponding linear program (\ref{eq.core-LP1}) must satisfy the equalities
$$
      x^*(S_k^\psi) = \sum_{i\in S_k} x^*_i = v(S_k^\psi)  \quad\mbox{for $k = 1,\ldots,n$},
$$
which means that $x^*$ is exactly the primal \textsc{Monge} vector $x^\pi$ and, hence, that $x^\pi\in
\core(v)$ holds.

\qed

\subsection{Concavity}\label{sec:core-concavity} Let us call the characteristic function $v:2^N\to\R$
\emph{concave} if $v$ arises
from the restriction of a concave  function to the $(0,1)$- in\-cidence vector $c^{(S)}$ of the coalitions
$S$, \emph{i.e.}, if if there is a concave function $f:\R^N\to \R$ such that
$$
   v(S) =  f(c^{(S)}) \quad\mbox{holds for all $S\subseteq N$.}
$$

Accordingly, the cooperative game $(N,v)$  is \emph{concave}\index{concave! cooperative game} if
$v$ is concave.  We will not pursue an investigation of general concave cooperative games here but focus
on a particularly important class of concave games which are closely tied to the \textsc{Monge} algorithm
{\it via} Theorem~\ref{t.Monge-LP}.

\medskip
\begin{tcolorbox}
\begin{proposition}\label{p.Monge-concave} If all \textsc{Monge} vectors of the game $(N,v)$ lie in
$\core(v)$, then $(N,v)$ is concave.
\end{proposition}
\end{tcolorbox}

\medskip
\begin{remark}\label{r.concave-games} The converse of Proposition~\ref{p.Monge-concave} is not  true:
there are concave games whose core does not include all primal \textsc{Monge} vectors.
\end{remark}

\medskip
\paragraph{\bf A word of terminological caution.}\index{game! convex}  The game-theoretic literature often
applies the termino\-logy "convex cooperative game'' to games $(N,v)$ having all primal \textsc{Monge}
vectors in $\core(v)$.  In our terminology, however, such games are not \emph{convex} but \emph{concave}.

\medskip
To avoid terminological confusion, one may prefer to refer to such games as   \emph{super\-modular games}
(\emph{cf.} Theorem~\ref{t.Monge-algorithm}  below).

\subsection{Supermodularity}\label{sec:supermodularity}

\subsection{Submodularity}\label{sec:submodularity} A characteristic function $v$ is called
\emph{submodular}\index{submodular} \index{game! submodular} if the inequality
$$
     v(S\cap T) +v(S\cup T) \leq v(S) +v(T)  \quad\mbox{holds for all $S,T\subseteq N$}.
$$

\medskip
\begin{ex}\label{ex.submodular} Show for the zero-normalized game $(N,v)$ the equivalence of the
statements:

\begin{enumerate}
\item $v$ is supermodular.
\item $v^*$ is submodular.
\item $w=-v$ is submodular.
\end{enumerate}
\end{ex}

\medskip
In view of the equality $\core(c^*) = \core^*(c)$ (Ex.~\ref{ex.profit-core}) ,  we find that the
\textsc{Monge} algorithm also constructs vectors in the $\core^*(c)$ of cooperative cost games $(N,c)$
with submodular characteristic functions $c$.

\medskip
\begin{remark}\label{r.Monge-network-vector} Note the fine point of Theorem~\ref{t.Monge-algorithm}, which
in the language of submodularity says: \emph{$(N,c)$ is a submodular cost game if and only if \emph{all}
\textsc{Monge}  vectors $x^\pi$ lie in $\core^*(c)$.}

\medskip
Network connection games are typically \emph{not} submodular. Yet, the particular greedy cost distribution vector
discussed in Section~3.3 does lie in $\core^*(c)$, as the ambitious reader is invited
to demonstrate.
\end{remark}

\medskip
\begin{remark}\label{r.discrete-optimization} Because of the \textsc{Monge}  algorithm, sub- and
supermodular functions play a prominent role in the field of discrete optimiziation\footnote{see,
\emph{e.g.},  \textsc{S. Fujishige}~\cite{Fujishige2005}}. In fact, many results of discrete optimization
have a direct interpretation in the theory of cooperative games. Conversely, the model of cooperative
games often provides conceptual insight into the structure of discrete optimization problems.
\end{remark}

\medskip
\begin{remark}[Greedy algorithm]\label{r.greedy-algorithm}\index{greedy!algorithm} The \emph{Monge}
algorithm, applied to linear programs with core-type constraints is also known as the \emph{greedy
algorithm} in discrete optimization.
\end{remark}

\section{Values}\label{sec:TU-values}\index{value}
While the marginal value $\partial_i v(S)$ of player $i$'s decision to join resp. to leave the coalition
$S$  is intuitively clear, it is less clear how the overall strength of $i$ in a game should be assessed.
From a mathematical point of view, there are infinitely many possibilities to do this.

\medskip
In general, we understand by a \emph{value} for the class of all TU-games $(N,v)$ a vector-valued function
$$
  \Phi:\R^\cN\to \R^N
$$
that associates with every characteristic function $v$ a vector $\Phi(v)\in \R^N$. Given $\Phi$, the
coordinate value $\Phi_i(v)$ is the assessment of the strength of $i\in N$ in the game $(N,v)$ according
to the evaluation concept $\Phi$.

\subsection{Linear values}\label{sec:linear-values}

\subsection{Random values}\label{sec:random-values}\index{value!random}

\subsubsection{\bf The value of \textsc{Banzhaf}} As an example, let us assume that a player $i$ joins any
of the $2^{n-1}$ coalitions $S\subseteq N\setminus\{i\}$ with equal likelihood, \emph{i.e.}, with
probability
$$
    \pi_S^B = \frac{1}{2^{n-1}}.
$$

Consider the unanimity game $v_T=\widehat{\delta}_T$ and observe that $\partial_i v_T(S) = 0$ holds if
$i\notin T$. On the other hand, if $i\in T$, then one has
$$
  \partial_iv_T(S) = 1 \quad \Longleftrightarrow \quad T\setminus \{i\} \subseteq S.
$$
So the number of coalitions $S$ with $\partial_iv_T(S)=1$ equals
$$
|\{S\subseteq N\setminus \{i\}\mid T\subseteq S\cup\{i\}\}| = 2^{n-|T|-1}.
$$

\medskip
Hence we conclude

\begin{equation}\label{eq.Banzhaf-probability}
E^{\pi^B}_i(v_T) = \sum_{S\subseteq N\setminus\{i\}} \partial_iv_T(S)\pi_S^B = \frac{2^{n-|T|-1}}{2^{n-1}}
= \frac{1}{2^{|T|}},
\end{equation}

which means that the random value $E^{\pi^B}$ is identical with the \textsc{Banzhaf} power index. The
probabilistic approach yields the explicit formula

\begin{equation}\label{eq.Banzhaf-formula}
 \Phi^B_i(v) = E^{\pi^B}_i(v) = \frac{1}{2^{n-1}}\sum_{S\subseteq N\setminus\{i\}}(v(S\cup i)-v(S))
 \quad(i\in N).
\end{equation}

\subsubsection{\bf Marginal vectors and the \textsc{Shapley} value}\label{sec:marginal-vector}

\section{Boltzmann values}\label{sec:Boltzmann-values} \index{value!\textsc{Boltzmann}}
The probabilistic analysis of the previous section shows that the value assessment concepts of the
\textsc{Banzhaf} power index and the \textsc{Shapley} value, for example, implicitly assume that players
just join -- but never leave -- an existing coalition in a cooperative game $(N,v)$.

\medskip
In contrast, the  model of the present section assumes an underlying  probability distribution $\pi$ on
the set $2^N$ of \emph{all} coalitions of  $N$  and assigns to player $i\in N$ its expected marginal value
$$
   E_i(v,\pi) = \sum_{S\subseteq N} \partial_iv(S)\pi_S.
$$

\medskip
\begin{ex}\label{ex.marginal-sum-uniform-distribution} Let $\pi$ be the uniform distribution on $\cN$:
$$
     \pi_S = \frac{1}{|\cN|} \quad\mbox{for all $S\in \cN$.}
$$
In view of
\begin{eqnarray*}
\sum_{S\subseteq N} \partial_iv(S) &=& \sum_{i\in S} v(S)-v(S\setminus i)) + \sum_{i\not\in
S}(v(S)-v(S\cup i))\\
&=& \sum_{T\subseteq N\setminus i} (v(T\cup i)-v(T))  +\sum_{T\subseteq N\setminus i}(v(T) - v(T\cup i))
\\
&=&0,
\end{eqnarray*}
one has
$$
E_i(v,\pi) = \sum_{S\in \cN}\partial_i v(S)\pi_S = \frac{1}{|\cN|}\sum_{S\in \cN}\partial_i v(S) = 0.
$$
So the expected marginal value of any particular  player is zero, if all co\-alitions are equally likely.

\end{ex}

\medskip

We furthermore do allow $\pi$ to depend on the particular characteristic function $v$ under consideration. It follows that the functional $v\mapsto E_i(v,\pi)$ is not guaranteed to be linear.

\medskip
The idea of the \textsc{Boltzmann} value is based on the fact that one expects the characteristic value
$$
     \mu =E(v,\pi) =\sum_{S\in \cN} v(S)\pi_S
$$
if the players agree on a  coalition $S\in \cN$ with probability $\pi_S$. So we may associate with $\mu$
its \textsc{Boltzmann} temperature $T$ and define the corresponding \textsc{Boltzmann}
\emph{values}\index{\textsc{Boltzmann}! value}
\begin{equation}\label{eq.Boltzmann-value}
E^T_i(v) = \frac{1}{Z_T}\sum_{S\in \cN} \partial_i v(S) e^{v(S)/T} \quad(\mbox{with $Z_T = \sum_{S\in \cN}
e^{v(S)/T}$})
\end{equation}
for the players $i$ in the cooperative game $(N,v)$ with expected characteristic value $\mu$.

\section{Coalition formation}\label{sec:coalition-formation}

\subsection{Individual greediness and public welfare}\label{sec:greediness-welfare} Let us assume that $N$
is a society whose common welfare is expressed by the potential $v$ on the family $\cN$ of all possible
coalitions: If the members of $N$ decide to join in a coalition $S\subseteq N$, then the value $v(S)$ will
be produced.

\medskip
If all members of $N$ act purely greedily, an $i\in N$ has an incentive to change its decision with
respect to the current coalition $S$ depending on its marginal value $\partial_i v(S)$ being positive or
negative. This behavior, however, will not guarantee a high public welfare.

\medskip
The \textsc{Metropolis} process suggests that the public welfare can be steered if an incentive is provided
such that $i$ enacts a move $S\to S\Delta\{i\}$ (\emph{i.e.}, changes its decision) with a non-zero
probability
$$
      \alpha^T_i(S) = e^{\partial_iv(S)/T} \quad\mbox{(even) if\; $\partial_iv(S) <0$.}
$$

If the control parameter $T>0$ is sufficiently small, the behavior of an $i\in N$ is ''almost purely
greedy'' in the sense
$$
T\to 0 \quad \Longrightarrow\quad \alpha_i^T(S) \to 0  \quad\mbox{if $\partial_iv(S) <0$.}
$$

\medskip
Moreover, a small temperature $T>0$ in the coalition formation process allows us to expect a high public
welfare. \index{welfare}

\subsection{Equilibria in cooperative games}\label{sec:cooperative-equilibria}
In many cooperative games, the grand coalition offers an obvious equilibrium if the players' utilities are
assessed by their marginal values:

\medskip
\begin{tcolorbox}
\begin{lemma}\label{l.grand-equilibrium} Let $(N,v)$ be a cooperative game. Then the two statements are
equivalent:
\begin{enumerate}
\item $N$ is a gain equilibrium with respect to the individual utility functions $u_i(S) =
    \partial_iv(S)$.
\item $v(N) \geq v(N\setminus i)$ for all $i\in N$.
\end{enumerate}

\end{lemma}
\end{tcolorbox}

\medskip
In general, we may view $(N,v)$ as a $n$-person matrix game with individual utilities
$$
     u_i(S) = \partial_i v(S) = v(S\Delta i) - v(S).
$$

Hence we know from \textsc{Nash}'s  Theorem~\ref{t.Nash} that the randomization of $(N,v)$ admits an
equilibrium.

\medskip
\begin{remark}\label{r.cooperative-randomization} The randomization of $(N,v)$ means that each $i\in N$
selects a probability $0\leq w_i\leq 1$ for the probability to become active. The coalition $S$ is thus
formed with probability
$$
    w(S) = \prod_{i\in S} w_i \prod_{j\notin S} (1-w_j).
$$
The expected value of $v$ is thus
$$
    E(v,w) = \sum_{S\subseteq N} v(S) w(S).
$$
The randomization of $(N,v)$ essentially is a fuzzy game (see Ex.~\ref{ex.fuzzy-games}) with
potential function
$$
       \ov{v}(v) = E(v,w) \quad(w\in [0,1]^N).
$$
\end{remark}

\medskip
Observe, in contrast to the above:

\medskip
\begin{tcolorbox}
\begin{itemize}
\item The \textsc{Boltzmann} coalition formation model does \emph{not} admit coalition equilibria at
    temperature $T\neq 0$, unless $v$ is constant,
\end{itemize}
\end{tcolorbox}

\medskip
but implies high public welfare if the temperature is small.

\medskip
Many value concepts (like \textsc{Shapley} and \textsc{Banzhaf}, for example), are based on marginal gains
with respect to having \emph{joined} a coalition as fundamental criteria for the individual utility
assessment of a player.

\medskip
So let us consider the cooperative game $(N,v)$ and take into account that the game will eventually split
$N$ into a group $S \subseteq N$ and the complementary group $S^c =N\setminus S$. Suppose a player $i\in
N$ evaluates its utility relative to the partition  $(S,S^c)$ of $N$ by
$$
     v_i(S) = v_i(S^c) = \left\{\begin{array}{cl} v(S)-v(S\setminus i) &\mbox{if $i\in S$}\\
                              v(S^c)-v(S^c\setminus i) &\mbox{if $i\in S^c$.}\end{array}\right.
$$

\medskip
\begin{ex}\label{ex.supermodular-N-equilibrium} Assume that $(N,v)$ is a supermodular game. Then one has
for all players $i\neq j$,
\begin{eqnarray*}
v_i(N) = v(N) -v(N\setminus i) &\geq& v(N\setminus j) - v((N\setminus j)\setminus i) =v_i(N\setminus j)\\
v_i(N) = v(N) -v(N\setminus i) &\geq& v(\{i\})-v(\emptyset) = v_i(N\setminus i).
\end{eqnarray*}
Consequently, the grand coalition $N$ represents a gain equilibrium relative to the utilities $v_i$.
\end{ex}

\medskip
\begin{ex}\label{ex.submodular-N-equilibrium} Assume that $(N,c)$ is a zero-normalized submodular game
and that the players $i$ have the utilities
$$
     c_i(S) = \left\{\begin{array}{cl} c(S)-c(S\setminus i) &\mbox{if $i\in S$}\\
                              c(S^c)-c(S^c\setminus i) &\mbox{if $i\in S^c$.}\end{array}\right.
$$
\emph{Show:} The grand coalition $N$ is a cost equilibrium relative to the utilities~$c_i$.
\end{ex}

\chapter{Interaction Systems and Quantum Models}\label{chap:Interaction-quantum-models}

\begin{tcolorbox}
This final chapter investigates game-theoretic systems making use of the algebra of complex numbers. Not only cooperation models are generalized, but also interaction of pairs on elements of a  set
$X$ finds an appropriate setting. The states are naturally represented as hermitian matrices with complex coefficients. This representation allows one to carry out standard spectral analysis for
interaction systems and provides a link to the corresponding mathematical model of quantum systems in physics.

\medskip
While the analysis could be extended to general \textsc{Hilbert} spaces, $X$ is assumed to be finite to
keep the discussion straigthforward.\footnote{see also \textsc{Faigle} and \textsc{Grabisch}~\cite{FaigleGrabisch2017}}

\medskip
It is historically perhaps surprising that \textsc{John von Neumann}, who laid
out the mathematical foundations of quantum theory\footnote{\textsc{von Neumann}~\cite{vNeumann2018}},
did not build game theory on the same mathematics in his work with \textsc{Oskar Morgenstern}.
\end{tcolorbox}

\section{Algebraic preliminaries}\label{sec:algebraic-preliminaries}  Since matrix algebra is the main
tool in our analysis, we review some more fundamental notions from linear algebra (and
Chapter~\ref{sec:mathematical-preliminaries}). Further details and proofs can be found in any decent book
on linear algebra\footnote{\emph{e.g.}, \textsc{Nering}~\cite{Nering1967}}.

\medskip
Where $X=\{x_1,\ldots,x_m\}$ and $Y=\{y_1,\ldots, y_n\}$  are two finite index sets, recall that
$\R^{X\times Y}$  denotes the real vector space of all matrices $A$ with rows indexed by $X$, columns
indexed by $Y$, and coefficients $A_{xy}\in \R$.

\medskip
The \emph{transpose} of $A\in \R^{X\times X}$ is the matrix $A^T\in \R^{Y\times X}$ with the coefficients
$A^T_{xy} = A_{xy}$. The map $A\mapsto A^T$  establishes an isomorphism between the vector spaces
$\R^{X\times Y}$ and $\R^{Y\times X}$ .

\medskip
Viewing $A\in \R^{X\times Y}$ and $B\in \R^{Y\times X}$ as $mn$-dimensional parameter vectors, we have the
usual euclidian inner product as
$$
  \(A|B\) = \sum_{(x,y)\in X\times Y}  A_{xy}B_{xy}  = \tr(B^TA),
$$
where $\tr C$ denotes the trace of a matrix  $C$. In the case $\(A|B\) = 0$, $A$ and $B$ are said to be \emph{orthogonal}.\index{orthogonal}   The
associated euclidian norm is \index{norm! euclidian}
 $$
 \|A\| =\sqrt{\(A|A^T\)} =\sqrt{\sum_{(x,y)\in X\times Y } |A_{xy}|^2}.
$$
We think of a vector $v\in \R^X$ typically as a column vector. $v^T$ is the row vector with the same
coordinates $v^T_x = v_x$. Be aware of the difference between the two matrix products:

\begin{eqnarray*}
v^Tv &=& \sum_{x\in X} |v_x|^2 = \|v\|^2  \\
vv^T &=& \begin{bmatrix} v_{x_1}v_{x_1} &v_{x_1}v_{x_2} &\ldots &v_{x_1}v_{x_m}\\
 v_{x_2}v_{x_1} &v_{x_2}v_{x_2} &\ldots &v_{x_1}v_{x_m}\\
 \vdots &\vdots &\ddots &\vdots \\
  v_{x_m}v_{x_1} &v_{x_m}v_{x_2} &\ldots &v_{x_m}v_{x_m}\end{bmatrix} .
\end{eqnarray*}

\subsection{Symmetry decomposition}\label{sec:symmetry-decomposition}
Assuming identical index sets \index{symmetry decomposition}
$$
X=Y = \{x_1,\ldots,x_n\}
$$
 a matrix $A\in \R^{X\times X}$ is \emph{symmetric} if $A^T = A$. In the case $A^T= -A$, the matrix $A$ is
 \emph{skew-symmetric}. With an arbitrary matrix $A\in \R^{X\times X}$, we associate the matrices
 $$
 A^+ =\frac12(A+A^T) \quad\mbox{and}\quad A^- = \frac12(A-A^T) = A - A^+.
 $$
Notice that $A^+$ is symmetric and $A^-$ is skew-symmetric. The \emph{symmetry decomposition} of $A$ is
the representation

\begin{equation}\label{eq.symmetry-decomposition}
\fbox{$\; A = A^+ +A^-\;$}
\end{equation}

The matrix $A$ allows exactly one decomposition into a symmetric and a skew-symmetric matrix (see
Ex.~\ref{ex.symmetry-decomposition}). So the symmetry decomposition is unique.

\medskip
\begin{tcolorbox}
\begin{ex}\label{ex.symmetry-decomposition} Let $A,B,C\in \R^{X\times X}$ be such that $A = B+C$. Show
that the two statements are equivalent:

\begin{enumerate}
\item $B$ is symmetric and $C$ is skew-symmetric.
\item $B=A^+$ and $C = A^-$.
\end{enumerate}
\end{ex}
\end{tcolorbox}

\medskip
Notice that symmetric and skew-symmetric matrices are necessarily pairwise orthogonal (see
Ex.\ref{ex.Pythagoras}).

\medskip
\begin{ex}\label{ex.Pythagoras} Let $A$ be a symmetric and $B$ a skew-symmetric matrix. Show:
$$
   \(A|B\) = 0 \quad\mbox{and}\quad \|A+B\|^2 =  \|A\|^2 + \|B\|^2.
$$
\end{ex}

\section{Complex matrices}\label{sec:complex-matrices}
In physics and engineering, complex numbers offer a convenient means to represent orthogonal structures.
Applying this idea to the symmetry decomposition, one arrives at so-called \emph{hermitian matrices}.

\medskip
\paragraph{\bf Inner products.} Recall that a \emph{complex number} $z\in \C$ is an expression of the form $z=a+\im b$ where $a$ and $b$
are real numbers and $\im$ a special ''new'' number, the \emph{imaginary unit}, with the property $\im^2
=-1$. The squared absolute value of the complex number $z=a+\im b$ is
$$
    |z|^2 = a^2 +b^2 = (a-\im b)(a+\im b) = \ov{z}z,
$$
with $\ov{z} = a-\im b$ being the \emph{conjugate} of $z$. More generally, we define the \emph{hermitian product}\index{product! hermitian} of two complex numbers $u,v\in \C$ as the complex number
\begin{equation}\label{eq.complex-hermitian-product}
\(u|v\) = \ov{v} u.
\end{equation}

The \emph{(hermitian) inner product} of two vectors $u,v\in \C^X$ with components $u_x$ and $v_x$ is the complex number
$$
 \(u|v\) = \sum_{x\in X} \(u_x|v_x\).
$$
The \emph{length} (or \emph{norm}) \index{norm! complex} of a vector $u = a +\im b \in \C^X$ (with $a,b\in \R^X$) is
$$
\|u\| = \sqrt{\sum_{x\in X} \(u_x|u_x\)} =\sqrt{\sum_{x\in X} |u_x|^2} = \sqrt{\sum_{x\in X} |a_x|^2 + |b_x|^2}.
$$

\medskip
\paragraph{\bf Conjugates and adjoints.} The \emph{conjugate} of a vector $v\in \C^X$ is the vector $\ov{v}\in \C^X$ with the conjugated components $\ov{v}_x$.  The vector $v^* =\ov{v}^T$ is the \emph{adjoint} \index{adjoint} of $v$. With this notation, the inner product \index{inner product} of the column vectors $u,v\in \C^X$ is
$$
\(u|v\) = \sum_{x\in X} u_x v^*_x = v^*u,
$$
where we think of the $1\times 1$ matrix $v^*u$ just as a complex number. Accordingly, the adjoint of the matrix $C\in \C^{X\times Y}$ is the matrix
$$
    C^* = \ov{C}^T \in \C^{Y\times X}.
$$

\medskip
\begin{ex}[Trace]\label{ex.complex-trace} The inner product of the matrices  $U,V\in \C^{X\times Y}$ is
$$
\(U|V\) = \tr(V^*U).
$$
\end{ex}

\medskip
The matrix $C\in \C^{X\times X}$ is \emph{selfadjoint} \index{selfadjoint} if it equals its adjoint, \emph{i.e.}, if
$$
       C = C^* = \ov{C}^T.
$$

\medskip
\begin{ex} Let $v\in \C^X$ be a column vector. Then $vv^*\in \C^{X\times X}$ is a selfadjoint matrix of norm $\|vv^*\| =\|v\|^2$.
\end{ex}

\subsection{Spectral decomposition}\label{sec:selfadjointness}
If (and only if) the matrix $C\in \C^{X\times X}$ has real coefficients,
$$\ov{C}~=~ C
$$
holds and the notion 'selfadjoint' boils down to 'symmetric'. It is well-known that real symmetric matrices can be diagonalized. With the same technique,
one can extend the diagonalization to general selfadjoint matrices:

\medskip\index{spectral!theorem}
\begin{tcolorbox}
\begin{theorem}[Spectral Theorem]\label{t.spectral-theorem}   For a matrix $C\in \C^{X\times X}$ the two
statements are equivalent:
\begin{enumerate}
\item $C=C^*$.
\item $\C^X$ admits a unitary basis $U=\{U_x\mid x\in X\}$ of eigen\-vectors $U_x$ of $C$ with real
    eigenvalues $\lambda_x$.
\end{enumerate}
\end{theorem}
\end{tcolorbox}

\medskip\index{unitary}\index{eigenvalue}
\emph{Unitary} means for the basis $U$ that the vectors $U_x$ have unit norm and are pairwise orthogonal in the sense
$$
\(U_x|U_y\) = \left\{\begin{array}{cl} 1 &\mbox{if $x=y$}\\ 0 &\mbox{if $x\neq
y$.}\end{array}\right.
$$
The scalar $\lambda_x$ is the \emph{eigenvalue} of the \emph{eigenvector} $U_x$ of $C$ if
$$
      C U_x = \lambda_x U_x.
$$

\medskip
It follows from Theorem~\ref{t.spectral-theorem} (see Ex.~\ref{ex.spectral-decomposition}) that a
selfadjoint matrix $C$ admits a \emph{spectral\footnote{the \emph{spectrum} of a matrix is, by definition,
its set of eigenvalues} decomposition}, \emph{i.e.}, a representation in the form

\begin{tcolorbox}\index{spectral decomposition}
\begin{equation}\label{eq.spectral-decomposition}
 C = \sum_{x\in X} \lambda_x U_xU_x^*,
 \end{equation}
 where the $U_x$ are pairwise orthogonal eigenvectors of $C$ with eigenvalues $\lambda_x\in \R$.
\end{tcolorbox}

 \medskip
 \begin{ex}\label{ex.spectral-decomposition}  Let $U=\{U_x\mid x\in X\}$ be a unitary basis of $\C^X$
 together with a set $\Lambda = \{\lambda_x\mid x\in X\}$ a set of arbitrary complex scalars. Show:
 \begin{enumerate}
 \item The $U_x$ are eigenvectors with eigenvalues $\lambda_x$ of the matrix
 $$
      C = \sum_{x\in X} \lambda_x U_xU_x^*.
 $$
 \item $C$ is selfadjoint if and only if all the $\lambda_x$ are real numbers.
 \end{enumerate}
\end{ex}

\medskip
The spectral decomposition shows:

\medskip
\begin{tcolorbox}
\emph{The selfadjoint matrices $C$ in $\C^{X\times X}$ are precisely the linear com\-binations of matrices
of type
$$
C = \sum_{x\in X}\lambda_x U_xU_x^*,
$$
where the  $U_x$ are (column) vectors in $\C^X$ and the $\lambda_x$ are real numbers.}
\end{tcolorbox}

\medskip
\paragraph{\bf Spectral unity decomposition.}  As an illustration, consider a unitary matrix $U\in \C^{X\times X}$, \emph{i.e.} a matrix  with pairwise orthogonal column vectors $U_x$ of norm $\|U\|_x=1$, which means that the identity matrix
$I$ has the representation
$$
                I = UU^* =U^*U.
$$
The eigenvalues of $I$ have all value $\lambda_x =1$. Relative to $U$, the matrix $I$ has the spectral
decomposition
\begin{equation}\label{eq.unity-decomposition}
I = \sum_{x\in X} U_xU_x^*.
\end{equation}

\medskip
For any vector $v\in \C^X$ with norm $\|v\| =1$, we therefore find
\begin{eqnarray*}
  1 = \(v|v\) = v^*Iv &=&  \sum_{x\in X} v^*U_xU_x^*v \\
&=& \sum_{x\in X} \ov{\(v|U_x\)}\(v|U_x\) = \sum_{x\in X} |\(v|U_x\)|^2.
\end{eqnarray*}
It follows that the (squared) absolute values
$$
p_x^v = |\(v|U_x\)|^2 \quad(x\in X)
$$
yield the components of a probability distribution $p^v$ on the set $X$. More generally, if the selfadjoint matrix $C$ with eigenvalues $\rho_x$ has the form
$$
C = \sum_{x\in X} \rho_x U_xU^*_x ,
$$
then we have for any $v\in \C^X$,

\begin{equation}\label{eq.eigenvalue-expectation}
\(v|Cv\)  = v^*Cv = \sum_{x\in X} \rho_x |\(v|U_x\)|^2  = \sum_{x\in X} \rho_x p_x^v.
\end{equation}

In other words:

\medskip
\begin{tcolorbox}
\emph{The inner product $\(v|Cv\)$ of the vectors $v$ and $Cv$ is the expected value of the eigenvalues
$\rho_x$ of $C$ with respect to the probability distribution $p^v$ on $X$.}
\end{tcolorbox}

\medskip
\begin{ex}[Standard unity decomposition]\label{ex.standard-unity-decomposition} The unit vectors $e_x\in \C^X$ yield the standard unity decomposition
$$
    I = \sum_{x\in X} e_x e_x^*.
$$
Accordingly, a vector $v\in \C^X $ of length $\|v\|=1$ with the components $v_x$ implies the standard probability distribution on $X$ with the components
$$
p^v_x = |\(v|e_x\)|^2 = |v_x|^2.
$$
\end{ex}

\subsection{Hermitian representation}\label{sec:hermitian-representation}\index{hermitian! representation}
\index{hermitian! matrix} \index{matrix! hermitian}
Coming back to real matrices in the context of symmetry decompositions, let us associate with a real matrix $A\in
\R^{X\times X}$ the complex matrix
$$
\hat{A} = A^+ +\im A^-.
$$
$\hat{A}$ is a \emph{hermitian\footnote{\textsc{C. Hermite} (1822-1901)} matrix}. The \emph{hermitian map}
$A\mapsto \hat{A}$ establishes an iso\-morphism between the vector space $\R^{X\times X}$ and the vector
space
$$
      \H_X =\{\hat{A}\mid A\in \R^{X\times X}\}
$$
with the set $\R$ as field of scalars\footnote{$\H_X$ is not a complex vector space: The product $z C$ of
a hermitian matrix $C$ with a complex scalar $z$ is not necessarily hermitian.}. The import in our context
is the fundamental observation that the selfadjoint matrices are precisely the hermitian matrices:

\medskip
\begin{tcolorbox}
\begin{lemma}\label{l.Hermite} Let $C\in \C^{X\times X}$ be an arbitrary complex matrix. Then
$$
     C\in \H_X \quad \Longleftrightarrow \quad C=C^*
$$
\end{lemma}
\end{tcolorbox}

\Pf Assume $C= A+\im B$ with $A,B\in \R^{X\times}$ and hence
$$
   C^* = A^T -\im B^T
$$
So $C=C^*$  means  symmetry $A=A^T$ and skew-symmetry $B=-B^T$. Consequently, one has $\hat{A} = A$ and
$\hat{B}  = \im B$, which yields
$$
   C = A +\im B =\hat{A} + \hat{B}  \in \H_X.
$$
The converse is seen as easily.

\qed

\medskip
The remarkable property of the hermitian representation is:

\begin{tcolorbox}
\begin{itemize}
\item While a real matrix $A\in \R^{X\times X}$ does not necessarily admit a spectral decomposition
    with real eigenvalues, its hermitian representation $\hat{A}$ is always guaranteed to have one.
\end{itemize}
\end{tcolorbox}

\medskip
\begin{ex}[\textsc{Hilbert} space]\label{ex.Hilbert-space}  Let $A,B\in \R^{X\times X}$ be arbitrary real matrices. Show:
$$
\(A|B\) = \(\hat{A}|\hat{B}\),
$$
\emph{i.e.}, inner products (and hence norms) are preserved under the hermitian representation. This means that $\R^{X\times X}$ and $\H_X$ are not only isomorphic as real vector spaces but also as (real) \textsc{Hilbert} spaces.
\end{ex}

\section{Interaction systems}\label{sec:interaction-systems}

\subsection{Interaction states}\label{sec:interaction-states}

\subsection{Interaction potentials}\label{sec:interaction-potentials}
\subsection{Interaction in cooperative games}\label{sec:cooperative-interaction}

\subsection{Interaction in infinite sets}\label{sec:infinite-interaction} Much of the current interaction
ana\-lysis remains valid for infinite sets with some modifications.

\medskip
For example, we admit as descriptions of interaction states only those matrices $A\in \R^{X\times X}$ with
the property
\begin{enumerate}
\item[(H1)] ${\rm supp}(A) = \{(x,y)\in X\times X\mid A_{xy}\neq 0\}$ is finite or countably infinite.
\item[(H2)]  $\|A\|^2 = \D\sum_{x,y\in X} |A_{xy}|^2 = 1. $
\end{enumerate}

\medskip
If the conditions (H1) and (H2) are met, we factually represent interaction states in \textsc{Hilbert}
{spaces}. To keep things simple, however, we retain the finiteness property of the agent set $X$  in the
current text and refer the interested reader to the literature\footnote{\emph{e.g.},
\textsc{Halmos}~\cite{Halmos1951} or \textsc{Weidmann}~\cite{Weidmann1980}} for further details.

\section{Quantum systems}\label{sec:quantum-systems}\index{quantum system}\index{system!quantum}
Without going into the physics of quantum mechanics, let us quickly sketch the basic mathematical model
and then look at the relationship with the interaction model. In this context, we think of an
\emph{observable}\index{observable}  as a mechanism $\alpha$ that can be applied to a system $\mathfrak
S$,
$$
\fbox{$\; \; \sigma\;\;$}\:\rightsquigarrow \fbox{$\; \alpha\;$}
\;\longrightarrow\;\alpha(\sigma)
$$
with the interpretation:

\begin{tcolorbox}
\begin{itemize}
\item If $\mathfrak S$ is in the state $\sigma$, then $\alpha$ is expected to  produce a measure\-ment
    result  $\alpha(\sigma)$.
\end{itemize}
\end{tcolorbox}

\subsection{The quantum model}\label{sec:quantum-model}

\subsection{Evolutions of quantum systems}\label{sec:quantum-evolution}

\subsection{The quantum perspective on interaction}\label{sec:quantum-interaction}

\subsection{The quantum perspective on cooperation}\label{sec:quantum-cooperation}

\section{Quantum games}\label{sec:quantum-games}\index{quantum!game}\index{system!quantum}
A large part of the mathematical analysis of game-theoretic systems follows the guideline

\begin{tcolorbox}
\begin{itemize}
\item \emph{Represent the system in a mathematical structure, analyze the representation mathematically
    and re-interpret the result in the original game-theoretic setting.}
\end{itemize}
\end{tcolorbox}

\medskip
When one chooses a representation of the system in the same space as the ones usually employed for the
representation of a quantum system, one automatically arrives at a ''quantum game'', \emph{i.e.}, at a
quantum-theoretic interpretation of a game-theoretic environment.

\medskip
So we understand by a \emph{quantum game} any game on a system $\mathfrak S$ whose states are represented
as quantum states and leave it to the reader to review game theory in this more comprehensive context.

\section{Final Remarks}\label{sec:final-remarks}
Why should one pass to complex numbers and the hermitian space $\H_X$ rather than the euclidian space
$\R^{X\times X}$ if both spaces are isomorphic real Hilbert spaces?

\medskip
The advantage lies in the algebraic structure of the field $\C$ of complex numbers, which yields the
spectral decomposition (\ref{eq.spectral-decomposition}), for example. It would be not impossible, but
somewhat ''unnatural'' to translate this structural insight back into the environment $\R^{X\times X}$
without appeal to complex algebra.

\medskip
Another advantage becomes apparent when one studies evolutions of systems over time. In the classical
situation of real vector spaces, \textsc{Markov} chains are important models for system evolutions. It turns out
that this model generalizes considerably when one passes to the context of \textsc{Hilbert}
spaces\footnote{\textsc{Faigle} and \textsc{Gierz}~\cite{FaigleGierz2017}}.

\medskip
The game-theoretic ramifications of this approach are to a large extent unexplored at this point.

\appendix
\chapter*{}\label{Appendix}
\setcounter{chapter}{1}\setcounter{section}{0}
\centerline{\bf \large Appendix}

\hspace{0.5cm}

\section{Basic facts from real analysis}\label{asec:basic-analysis}
(More details can be found in the standard literature~\footnote{\emph{e.g.},
\textsc{Rudin}~\cite{Rudin1953}}.) The \emph{euclidian norm} (or \emph{geometric length}) of a vector
$x\in \R^n$ with com\-ponents $x_j$, is
$$
\|x\| = \sqrt{x_1^2 +\ldots+ x_n^2}.
$$
The \emph{ball} with center $x$ and radius $r$ is the set
$$
B_r(x) = \{y\in \R^n\mid \|x-y\| \leq r\}.
$$

A subset $S\subseteq \R^n$ is \emph{closed}\index{closed} if each convergent sequence of elements $x_k\in S$ has its the limit point $x$ is also in $S$:
$$
    x_k \to x \quad \Longrightarrow \quad x\in S.
$$
The set $S$ is \emph{open}\index{open} if its complement $\R^n\setminus S$ is closed. The following statements are equivalent:
\begin{enumerate}
\item[(O')] $S$ is open.
\item[(O'')] For each $x\in S$ there is some $r>0$ such that $B_r(x)\subseteq S$.
\end{enumerate}

\medskip
The set $S$ is \emph{bounded}\index{bounded} if $S\subseteq B_r(0)$ holds for some $r\geq 0$. $S$ is said to be \emph{compact} \index{compact}  if $S$ is bounded and closed.

\medskip
\begin{lemma}[\textsc{Heine-Borel}]\label{al.Heine-Borel} $S\subseteq \R^n$ is compact if and only if
\begin{itemize}
\item[(HB)] every family $\cO$ of open sets $O\subseteq \R^n$ such that every $x\in S$ lies in at least
    one $O\in \cO$, admits a \emph{finite} number of sets $O_1,\ldots, O_\ell\in \cO$ with the covering
    property
\begin{itemize}
\item[$\bullet$] $S\subseteq O_1\cup O_2\cup\ldots \cup O_\ell$.
\end{itemize}
\end{itemize}
\qed
\end{lemma}

\medskip
It is important to note that compactness is preserved under forming direct products:
\begin{itemize}
\item If $X\subseteq \R^n$ and $Y\subseteq \R^m $ are compact sets, then $X\times Y\subseteq \R^{n+m}$
    is compact.
\end{itemize}

\medskip
\paragraph{\bf Continuity.} A function $f:S\to \R^m$ is \emph{continuous} if for all convergent sequences of elements $x_k\in S$, the sequence of function values $f(x_k)$ converges to the value of the limit:
$$
  x_k \to x \quad\Longrightarrow \quad f(x_k) \to f(x).
$$
The following statements are equivalent:
\begin{enumerate}
\item[(C')] $f:S\to \R^m$ is continuous.
\item[(C'')] For each open set $O\subseteq \R^m$, there exists an open set $O'\subseteq \R^n$ such that
$$
    f^{-1}(O) = \{x\in S\mid f(x)\in O\} = O'\subseteq S,
$$
\emph{i.e.}, the inverse image $f^{-1}(O)$ is open \emph{relative to $S$}.
\item[(C''')] For each closed set $C\subseteq \R^m$, there exists a closed set $C'\subseteq \R^n$ such that
$$
    f^{-1}(C) = \{x\in S\mid f(x)\in C\} = C'\subseteq S,
$$
\emph{i.e.}, the inverse image $f^{-1}(C)$ is closed \emph{relative to $S$}.
\end{enumerate}

\begin{lemma}[Extreme values]\label{al.continuous-extrema} If the real-valued function $f:S\to \R$ is continuous on the non-empty compact set $S\subseteq \R^n$,
then there exist elements $x_*, x^*\in S$ such that
$$
      f(x_*) \leq f(x) \leq f(x^*) \quad\mbox{holds for all $x\in S$.}
$$
\qed
\end{lemma}

\medskip
\paragraph{\bf Differentiability.}  The function $f:S\to \R$ is \emph{differentiable} on the open set
$S\subseteq  \R^n$ if for each $x\in S$ there is a (row) vector $\nabla f(x)$ such that for every
$d\in\R^n$ of unit length $\|d\|=1$, one has
 $$
\lim_{t\to 0}\frac{f(x+td)-f(x)}{t}  = \lim_{t\to 0} \frac{\nabla f(x)d}{t} \quad(t\in \R).
$$
$\nabla f(x)$ is the \emph{gradient} of $f$. Its components are the partial derivatives of $f$:
$$
 \nabla f(x) = \big(\partial f(x)/\partial x_1,\ldots, \partial f(x)/\partial x_n\big).
$$

\medskip
\textsc{Nota Bene.} All differentiable functions are continuous -- but not all continuous functions are
differentiable.

\section{Convexity}\label{asec:convex-sets}
A \emph{linear combination} of elements $x_1,\ldots,x_m$ is an expression of the form
$$
    z = \lambda_1 x_1 +\ldots +\lambda_m x_m,
$$
where $\lambda_1,\ldots,\lambda_m$ are scalars (real or complex numbers). The linear com\-bination $z$ with scalars $\lambda_i$ is \emph{affine} if
$$
   \lambda_1 +\ldots + \lambda_m = 1.
$$
An affine combination is a \emph{convex combination} if all scalars $\lambda_i$ are non\-negative real numbers. The vector $\lambda= (\lambda_1,\ldots,\lambda_m)$ of the $m$ scalars $\lambda_i$ of a convex combination is a  \emph{probability distribution} \index{probability distribution} on the index set
$$
X~=~\{1,\ldots,m\}.
$$

\medskip\index{convex! set}\index{convex! combination}
\paragraph{\bf Convex sets.}  The set $S\subseteq \R^n$ is \emph{convex} if it contains with every $x,y\in
S$ also the connecting line segment:
$$
 [x,y] = \{x+\lambda(y-x)\mid 0\leq \lambda \leq 1\} \subseteq S.
$$
It is easy to verify:
\begin{itemize}
\item The intersection of convex sets yields a convex set.
\item The direct product $S= X\times Y\subseteq \R^{n\times m}$ of convex sets $X\subseteq \R^n$ and
    $Y\subseteq \R^m$ is a convex set.
\end{itemize}

\medskip
\begin{ex}[Probability distributions]\label{aex.probability-distributions} For $X=\{1,\ldots,n\}$, the set
$$
     \cP(X) = \{(x_1,\ldots,x_n)\in \R^n\mid x_i\geq 0, x_1+\ldots +x_n = 1\}
$$
of all probability distributions on $X$ is convex. Because the function
$$
     f(x_1,\ldots,x_n) = x_1+ \ldots +x_n
$$
is continuous, the set
$$
   f^{-1}(1) = \{(x_1,\ldots,x_n)\in \R^n\mid f(x_1,\ldots,x_n) = 1\}
$$
is closed. The collection $\R^n_+$ of nonnegative vectors is closed in $\R^n$. Since the intersection of closed sets is closed, one deduces that
$$
\cP(X) = f^{-1}(1) \cap \R_+^n \subseteq B_n(0).
$$
is closed and bounded and thus compact.

\end{ex}

\medskip
\paragraph{\bf Convex functions.} \index{covex! function}
A function $f:S\to \R$ is \emph{convex (up)} on the convex set $S$ if for all $x,y\in S$  and for all
scalars $0\leq \lambda \leq 1$,
$$
    f(x+\lambda(y-x)) \geq f(x) +\lambda(f(y)-f(x))).
$$
This definition is equivalent to the requirement that one has for any finitely many  elements $x_1,\ldots,
x_m\in S$ and probability distributions  $(\lambda_1,\ldots,\lambda_m)$,
$$
f(\lambda_1x_1 +\ldots +\lambda_mx_m) \geq \lambda_1f(x_1) +\ldots +\lambda_m f(x_m).
$$
The function $f$ is \emph{concave} (or \emph{convex down}) if $g = -f$ is convex (up).

\medskip
A differentiable function $f:S\to \R$ on the open set $S\subseteq \R^n$ is convex (up) if and only if

\begin{equation}\label{aeq.diff-convex}
    f(y) \geq f(x) + \nabla f(x)(y-x) \quad\mbox{holds for all $x,y\in S$.}
\end{equation}

Assume, for example, that $\nabla f(x)(y-x)\geq 0$ it true for all $y\in S$, then one has
$$
      f(x) = \min_{y\in S} f(y).
$$

On the other hand, if $\nabla f(x)(y-x) < 0$ is true for some $y\in S$, one can move from $x$ a bit into
the direction of $y$ and find an element $x'$ with $f(x') <  f(x)$. Hence one has a criterion for
minimizers of $f$ on $S$:

\begin{lemma}\label{al.convex-minimum} If $f$ is a differentiable convex function on the convex set $S$,
then for any $x\in S$, the statements are equivalent:
\begin{enumerate}
\item $f(x) = \D\min_{y\in S} f(y)$.
\item $\nabla f(x)(y-x) \geq 0$ for all $y\in S$.
\end{enumerate}
\end{lemma}

\medskip
If strict inequality holds in (\ref{aeq.diff-convex}) for all $y\neq x$, $f$ is said to be \emph{stricly
convex}.

\medskip
In the case $n=1$ (\emph{i.e.}, $S\subseteq \R$), a simple criterion applies to twice differentiable
functions:
$$
\mbox{$f$ is convex} \quad \Longleftrightarrow \quad f''(x) \geq 0 \quad\mbox{for all $x\in S$.}
$$

For example, the logarithm function $f(x)=\ln x$ is seen to be strictly concave on the open interval $S=
(0,\infty)$ because of
$$
      f''(x) = -1/x^2 < 0 \quad\mbox{for all $x\in S$.}
$$

\section{Polyhedra and linear inequalities}\label{sec:linear-inequalities} \index{polyhedron}
A \emph{polyhedron} is the solution set of a finite system of linear equalities and inequalities. More precisely, if $A\subseteq \R^{m\times n}$ is a matrix and $b\in \R^m$ a parameter vector, then the (possibly empty) set
$$
P(A,b) = \{x\in \R^n\mid Ax\leq b\}
$$
is a \emph{polyhedron}. Since the function $x\mapsto Ax$ is linear (and hence continuous), one immediately checks that $P(A,b)$ is a closed convex subset of $\R^n$. Often, nonnegative solutions are of interest and one considers the associated polyhedron
$$
P_+(A,b) = \{x\in \R^n_+\mid Ax\leq b\} =\{x\in \R^n\mid Ax\leq b,-Ix\leq 0\},
$$
with the identity matrix $I\in \R^{n\times n}$.

\medskip
\begin{ex} The set $\cP(X)$ of all probability distributions on the finite set $X$ is a polyhedron. (\emph{cf.} Ex.~\ref{aex.probability-distributions}.)
\end{ex}

\medskip\index{\textsc{Farkas} lemma}
\begin{lemma}[\textsc{Farkas}\footnote{\textsc{Gy. Farkas (1847-1930)}}]\label{al.Farkas} If $P(A,b)\neq \emptyset$, then for any $c\in \R^n$ and $z\in \R$, the following statements are equivalent:
\begin{enumerate}
\item $c^Tx \leq z$ holds for all $x\in P(A,b)$.
\item There exists some $y\geq 0$ such that $y^TA = c^T$ and $y^Tb \leq z$.
\end{enumerate}
\end{lemma}

\medskip
Lemma~\ref{al.Farkas} is a direct consequence of the algorithm of \textsc{Fourier}\footnote{\textsc{J. Fourier (1768-1830)}}, which generalizes the Gaussian elimination method from systems of linear equalities to general linear inequalities\footnote{see, \emph{e.g.}, Section~2.4 in \textsc{Faigle} {\it et al.}~\cite{FaigleKernStill2002}}.

\medskip
The formulation of Lemma~\ref{al.Farkas} is one version of several equi\-valent characterizations of the solvability of finite linear (in)equality systems, known under the comprehensive label \emph{\textsc{Farkas} Lemma}. A nonnegative version of the \textsc{Farkas} Lemma is:

\medskip
\begin{lemma}[\textsc{Farkas}+]\label{al.Farkas+} If $P_+(A,b)\neq \emptyset$, then for any $c\in \R^n$ and $z\in \R$, the following statements are equivalent:
\begin{enumerate}
\item $c^Tx \leq z$ holds for all $x\in P_+(A,b)$.
\item There exists some $y\geq 0$ such that $y^TA \geq c^T$ and $y^Tb \leq z$.
\end{enumerate}
\end{lemma}

\medskip
\begin{ex}\label{aex.Farkas} Show that Lemma~\ref{al.Farkas} and Lemma~\ref{al.Farkas+} are equivalent. \emph{Hint:} Every $x\in \R^n$ is a difference $x=x^+ -x^-$ of two nonnegative vectors $x^+,x^-$. So
$$
Ax\leq b \quad\longleftrightarrow\quad  Ax^+ -Ax^- \leq b.
$$
\end{ex}

\section{\textsc{Brouwer}'s fixed-point theorem}\label{app:Brouwer-fixpoint} \index{fixed-point} A
\emph{fixed-point}\index{fixedpoint} of a map $f:X\to X$ is a point $x\in X$ such that $f(x) =x$. It is
usually difficult to find a fix-point (or even to decide whether a fixed-point exists). Well-known
sufficient conditions were formulated by \textsc{Brouwer}\footnote{\textsc{L.E.J. Brouwer} (1881-1966)}:

\medskip\index{fixed point}\index{\textsc{Brouwer}'s theorem}
\begin{theorem}[\textsc{Brouwer}]\label{at.Brouwer} Let $X\subseteq \R^n$ be a convex, compact and
non-empty set and $f:X\to X$ a continuous function. Then $f$ has a fixed-point.
\end{theorem}

\Pf See, \emph{e.g.}, the enyclopedic text of \textsc{Granas} and
\textsc{Dugundji}~\cite{GranasDugundji2003}.

\qed

\medskip
For game-theoretic applications, the following implication is of interest.

\medskip
\begin{corollary}\label{ac.concave-dominance} Let $X\subseteq \R^n$ be a convex, compact and nonempty set
and  $G:X\times X\to \R$  a continuous map that is concave in the second variable, \emph{i.e.},
\begin{itemize}
\item[(C)] for every $x\in X$, the map $y\mapsto G(x,y)$ is concave.
\end{itemize}
Then there exists a point $x^*\in X$ such that for all $y\in X$, one has
$$
    G(x^*,x^*) \geq G(x^*,y) \,.
$$
\end{corollary}

\Pf One derives a contradiction from the supposition that the Corollary is false. Indeed, if there is
no $x^*$ with the claimed property, then each $x\in X$ lies in at least one of the sets
$$
  O(y) = \{x\in X\mid G(x,x) < G(x,y)\} \quad(y\in X).
$$
Since $G$ is continuous, the sets $O(y)$ are open. Hence, since $X$ is compact, already finitely many
cover all of $X$, say
$$
X \subseteq O(y_1)\cup O(y_2)\cup \ldots \cup O(y_h).
$$
For all $x\in X$, define the parameters
$$
  d_\ell (x) = \max\{0, G(x,y_\ell) - G(x,x)\} \quad(\ell = 1,\ldots,h).
$$
$x$ lies in at least one of the sets $O(y_\ell)$. Therefore, we have
$$
   d(x) = d_1(x) + d_2(x) + \ldots +d_h(x) >0.
$$
Consider now the function
$$
x\mapsto \varphi(x) = \sum_{\ell=1}^j \lambda_\ell(x) y_i \quad\mbox{(with $\lambda_\ell =
d_\ell(x)/d(x)$).}
$$
Since $G$ is continuous, also the functions $x\mapsto d_\ell(x)$ are continuous. Therefore, $\varphi:X\to
X$ is continuous. By \textsc{Brouwer}'s Theorem~\ref{at.Brouwer}, $\varphi$ has a fixed point
$$
    x^* = \varphi(x^*) = \sum_{\ell =1}^h \lambda_\ell(x^*)y_\ell.
$$
Since $G(x,y)$ is concave in $y$ and $x^*$ is an affine linear combination of the $y_\ell$, we have
$$
G(x^*,x^*) = G(x^*,\varphi(x^*)) \geq \sum_{\ell=1}^h \lambda_\ell(x^*) G(x^*,y_\ell).
$$
If the Corollary were false, one would have
$$
\lambda_\ell G(x^*,y_\ell) \geq \lambda_\ell(x^*) G(x^*,x^*)
$$
for each summand and, in at least one case, even a strict inequality
$$
\lambda_\ell(x^*) G(x^*,y_\ell) > \lambda_\ell G(x^*,x^*),
$$
which would produce the contradictory statement
$$
G(x^*,x^*) > \sum_{\ell=1}^h \lambda_\ell(x^*) G(x^*,x^*) = G(x^*,x^*).
$$
It follows that the Corollary must be correct.
\qed

\section{The \textsc{Monge} algorithm}\label{asec:Monge-algorithm}\index{algorithm! \textsc{Monge}}
The \textsc{Monge}  \emph{algorithm}\index{\textsc{Monge}! algorithm} with respect to coefficient vectors $c,v\in \R^n$ has two versions.

\medskip
The \emph{primal} \textsc{Monge}  algorithm constructs a vector $x(v)$ with the components
$$
 x_1(v) = v_1 \quad\mbox{and}\quad   x_k(v) = v_k-v_{k-1} \quad(k=2,3,\ldots,n).
$$
The \emph{dual} \textsc{Monge}  algorithm constructs a vector $y(c)$ with the components
$$
   y_n (c) = c_n \quad\mbox{and}\quad y_\ell(c) = c_\ell - c_{\ell+1} \quad(\ell = 1,\ldots,n-1).
$$
Notice:
\begin{eqnarray*}
c_1\geq c_2 \geq \ldots\geq c_n &\Longrightarrow& y_\ell(c) \geq 0 \quad(\ell=1,\ldots,n-1)\\
v_1\leq v_2 \leq \ldots\leq v_n &\Longrightarrow& x_k(v) \geq 0 \quad(k=2,\ldots,n).
\end{eqnarray*}

\medskip
The important property to observe is

\medskip
\begin{lemma}\label{al.Monge-algorithm} The \textsc{Monge}  vectors $x(v)$ and $y(c)$ yield the equality
$$
    c^Tx(v) =  \sum_{k=1}^n c_k x_k(v)  = \sum_{\ell=1}^n v_\ell y_\ell (c) = v^Ty(c).
$$
\end{lemma}

\Pf Writing $x = x(v)$ and $y=y(c)$, notice for all $1\leq  k,\ell  \leq n$,
$$
    x_1+x_2+ \ldots+x_\ell= v_\ell  \quad\mbox{and}\quad y_k+y_{k+1}+\ldots+ y_n = c_k
$$
and hence
\begin{eqnarray*}
\sum_{k=1}^nc_k x_k = \sum_{k=1}^n\sum_{\ell=k}^n y_\ell  x_k = \sum_{\ell=1}^n\sum_{k=1}^\ell  x_ky_\ell
= \sum_{\ell=1}^n v_\ell y_\ell .
\end{eqnarray*}

\qed

\section{Entropy and \textsc{Boltzmann} distributions}\label{app:Boltzmann-distributions}
\subsection{\textsc{Boltzmann} distributions}\label{asec:Boltzmann-distributions}
The \emph{partition function}\index{partition function} $Z$ for a given vector $v=(v_1,\ldots,v_n)$ of
real numbers $v_j$ takes the (strictly positive) values \index{\textsc{Boltzmann}! distribution}
$$
   Z(t) = \sum_{j=1}^n e^{v_jt} \quad (t\in \R).
$$
The associated \textsc{Boltzmann} probability distribution $b(t)$ has the com\-ponents
$$
    b_j(t) = e^{v_jt}/Z(t) > 0
$$
and yields the expected value function
$$
\mu(t) = \sum_{j=1}^n v_jb_j(t) = \frac{Z'(t)}{Z(t)}.
$$
The \emph{variance} of $v$ is its expected quadratic deviation from $\mu(t)$:
\begin{eqnarray*}
  \sigma^2(t) &=& \sum_{j=1}^n (\mu(t)-v_j)^2 b_j(t) = \sum_{j=1}^n v_j^2 b_j(t) - \mu^2(t) \\
  &=& \frac{Z''(t)}{Z(t)} -\frac{Z'(t)^2}{Z(t)^2} =\mu'(t).
\end{eqnarray*}

One has $\sigma^2(t) \neq 0$ unless all $v_j$ are equal to a constant $K$ (and hence $\mu(t) = K$ for all
$t$).  Because $\mu'(t) = \sigma^2(t)>0 $, one concludes that $\mu(t)$ is strictly increasing in $t$ unless $\mu(t)$ is constant.

\medskip
Arrange the components of $v$ such that  $v_1\leq v_2\leq \ldots\leq v_n$. Then
$$
\lim_{t\to\infty} \frac{b_j(t)}{b_n(t)} =\lim_{t\to \infty} e^{(v_j -v_n)t} = 0 \quad\mbox{unless}\quad
v_j = v_n,
$$
which implies $b_j(t) \to 0$ if $v_j < v_n$. It follows that the limit distribution $b(\infty)$ is the
uniform distribution on the maximizers of $v$.
Similarly, one has
$$
\lim_{t\to -\infty} \frac{b_j(t)}{b_1(t)} =\lim_{t\to -\infty} e^{(v_j -v_1)t} = 0 \quad\mbox{unless}\quad
v_j = v_1
$$
and concludes that the limit distribution $b(-\infty)$ is the uniform distribution on the minimizers of
$v$.

\begin{theorem}\label{at.Boltzmann} For every value $v_1 < \xi < v_n$, there is a unique parameter
$t$ such that
$$
    \xi = \mu(t) = \sum_{j=1}^n v_j b_j(t).
$$
\end{theorem}

\Pf  The expected value function $\mu(t)$ is strictly monotone and continuous on $\R$ and satisfies
$$
        \lim_{\lambda\to -\infty} \mu(\lambda) =   v_1 \;\leq \mu(t) \;\leq v_n =\lim_{\lambda\to +\infty} \mu(\lambda).
$$
So, for every prescribed value $\xi$ between the extrema $v_1$ and $v_n$, there must exist precisely one
$t$ with $\mu(t) = \xi$.

\qed

\subsection{Entropy}\label{asec:entropy} The real function $h(x) = x\ln x$ is defined for all  nonnegative
real numbers\footnote{with the understanding $\ln 0 =-\infty$ and $0\cdot\ln 0 =0$} and has the strictly
increasing derivative
$$
   h'(x) = 1+ \ln x.
$$
So $h$ is strictly convex and satisfies the inequality
$$
     h(y)-h(x) > h'(x)(y-x) \quad\mbox{for all non-negative $y\neq x$.}
$$

$h$ is extended to nonnegative real vectors $x=(x_1,\ldots,x_n)$ {\it via}
$$
h(x) = h(x_1,\ldots,x_n) = \sum_{j=1}^n x_j\ln x_j \;\;\big( = \sum_{j=1}^n h(x_j)\big).
$$
The strict convexity of $h$ becomes the inequality
$$
   h(y) - h(x) > \nabla h(x)(y-x),
$$
with the gradient
$$
\nabla h(x) = (h'(x_1),\ldots,h'(x_n)) =(1+ \ln x_1,\ldots,1+\ln x_n).
$$

In the case $x_1+\ldots+x_n=1$, the nonnegative vector $x$ is a probability distribution on the set
$\{1,\ldots,n\}$ and has\footnote{by definition!}  the \emph{entropy}\index{entropy}
$$
   H(x) = \sum_{j=1}^n x_j\ln(1/x_j) = - \sum_{j=1}^n x_j\ln x_j  = -h(x_1,\ldots,x_n).
$$

We want to show that \textsc{Boltzmann} probability distributions are precisely the ones with maximal
entropy relative to given expected values.

\medskip
\begin{theorem}\label{at.max-entropy} Let $v=(v_1,\ldots,v_n)$ be a vector of real numbers and $b$ the
\textsc{Boltzmann} distribution on $\{1,\ldots,n\}$ with components
$$
    b_j = \frac{1}{Z(t)}e^{v_j t} \quad(j=1,\ldots,n).
$$
with respect to some $t$. Let $p=(p_1,\ldots,p_n)$ be a probability distribution with the same expected
value
$$
   \sum_{j=1}^n v_j p_j  =\mu = \sum_{j=1}^n v_jb_j.
$$
Then one has either $p =b$ or $H(p) < H(b)$.
\end{theorem}

\Pf For $d= p-b$, we have $\sum_j d_j = \sum_j p_j -\sum_j b_j = 1-1 =0$, and therefore
\begin{eqnarray*}
\nabla h(b)d &=& \sum_{j=1}^n d_j(1 + \ln b_j) \\
&=& \sum_{j=1}^n d_j v_jt +   (1 - \ln Z(t))\sum_{j=1}^n d_j = t \sum_{j=1}^n v_j d_j\\
&=& t\big(\sum_ {j=1}^n v_jp_j -\sum_{j=1}^n v_j b_j)\big) = t(\mu-\mu) = 0.
\end{eqnarray*}

In the case $p\neq b$, the strict convexity of $h$ thus yields
$$
   h(p) -h(b) > \nabla h(b)(p-b) = 0  \quad\mbox{and hence} \quad H(p) < H(b).
$$

\qed

\medskip\index{divergence}
\begin{lemma}[Divergence]\label{al.digergence} Let $a_1,\ldots,a_n, p_1,\ldots,p_n$ be arbitrary
non\-negative numbers. Then
$$
\sum_{i=1}^n a_i \leq \sum_{i=1}^n p_i\quad \Longrightarrow \quad \sum_{i=1}^np_i\ln a_i \leq \sum_{i=1}^n
p_i\ln p_i.
$$
Equality is attained exactly when $a_i=p_i$ holds for all $i=1,\ldots,n$.
\end{lemma}

\Pf We may assume $p_i\neq 0$ for all $i$ and make use of the well-known fact (which follows easily from
the concavity of the logarithm function):
$$
   \ln x\leq x-1 \quad\mbox{and}\quad \ln x = x-1 \Leftrightarrow x=1.
$$
Then we observe
$$
\sum_{i=1}^n p_i\ln\frac{a_i}{p_i} \leq \sum_{i=1}^n p_i(\frac{a_i}{p_i} -1) =\sum_{i=1}^n a_i -
\sum_{i=1}^n p_i \leq 0
$$
and therefore
$$
\sum_{i=1}^np_i\ln a_i - \sum_{i=1}^np_i\ln p_i = \sum_{i=1}^n p_i\ln\frac{a_i}{p_i}\leq 0.
$$
Equality can only hold if $\ln(a_i/p_i) = (a_i/p_i) -1$, and hence $a_i = p_i$ is true for all $i$.
\qed

\section{\textsc{Markov} chains}\label{asec:Markov-chains} A (\emph{discrete}) \textsc{Markov}
\emph{chain}\footnote{for more details, see, \emph{e.g.}, \textsc{Kemeny} and
\textsc{Snell}~\cite{KemenySnell1960}}  \index{\textsc{Markov}! chain} on a finite set $X$ is a (possibly
infinite) random walk on the graph $G=G(X)$ whose edges $(x,y)$ are labeled with probabilities $0\leq
p_{xy}\leq 1$ such that
$$
\sum_{y\in X} p_{xy} = 1 \quad\mbox{for all $x\in X$.}
$$
The walk starts in some node $s\in X$ and then iterates subsequent transitions $x\to y$ with probabilities
$p_{xy}$:
$$
   s \to x_1 \to x_2 \to \ldots \to x_n \to \ldots
$$

\medskip
Let $P=[p_{xy}]$ be the matrix of transition probabilities and $P^n=[p^{(n)}_{xy}]$ the $n$-fold matrix
product of $P$. Then the random walk has reached the node $y$ after $n$ iterations with probability
$$
     p^{(n)}_{sy}  \quad(y\in X).
$$
In other words: The $x_0$-row of $P^n$ is a probability distribution $p^{(n)}$ on $X$.

\medskip
The \textsc{Markov} chain is \emph{connected} if every node in $G$ can be reached from any other node with
non-zero probability in a finite number of transitional steps. This means:

\begin{itemize}
\item There exists some natural number $m$ such that $p^{(m)}_{xy} > 0$ holds for all $x,y\in X$.
\end{itemize}

\medskip
\begin{lemma}\label{al.Markov-convergence} If the \textsc{Markov} chain is connected and $p_{xx} >0$ holds
for at least one $x\in X$, then the \textsc{Markov} chain \emph{converges} in the following sense:
\begin{enumerate}
\item The limit matrix $P^\infty = \D\lim_{n\to\infty} P^n$ exists.
\item $P^\infty$ has identical row vectors $p^{(\infty)}$.
\item As a row vector, $p^{(\infty)}$ is the unique solution of the (in)equality system
$$
 p P = p  \quad\mbox{with $p_x\geq 0$ \;{and}\; $\D\sum_{x\in X} p_x = 1.$}
$$
\end{enumerate}

\qed
\end{lemma}

\medskip
A useful sufficient condition for the computation of a limit distribution $p^{(\infty)}$ of a
\textsc{Markov} chain is given in the next example.

\medskip
\begin{ex}\label{aex.Markov-convergence} Let $P=[p_{xy}]\in \R^{X\times X}$ be a \textsc{Markov}
transition probability matrix and $p\in \R^X$ a probability distribution on $X$ such that
$$
     p_x p_{xy} = p_yp_{xy} \quad\mbox{is true for all $x,y\in X$.}
$$
Then $P^T p = p$ holds. Indeed, one computes for each $x\in X$:
$$
\sum_{y\in X} p_y p_{xy} = \sum_{y\in X} p_x p_{xy} = p_x\sum_{y\in X}p_{yx} = p_x\cdot 1 = p_x.
$$
\end{ex}

\backmatter

\printindex

\end{document}